\documentclass[times,usenatbib,longnamesfirst]{mn2e}
\RequirePackage{graphicx}
\RequirePackage{subfigure}

\title[An observational study of the Ophiuchus cloud L1688]{The initial conditions of star formation VIII: an observational study of the Ophiuchus cloud L1688 and implications for the prestellar core mass function}

\author[Simpson, Nutter and Ward-Thompson]{R. J. Simpson,
D. Nutter and D. Ward-Thompson \\
School of Physics and Astronomy, Cardiff University,
The Parade, Cardiff, CF24 3AA}

\date{2008 July 8}

\pagerange{000--000}

\begin{document}

\label{firstpage}

\maketitle

\begin{abstract}
We re-analyse all of the archive observations of the Ophiuchus dark cloud L1688 that were carried out with the submillimetre common-user bolometer array (SCUBA) at the James Clerk Maxwell Telescope (JCMT). For the first time we put together all of the data that were taken of this cloud at different times to make a deeper map at 850$\mu$m than has ever previously been published. Using this new, deeper map we extract the pre-stellar cores from the data. We use updated values for the distance to the cloud complex, and also for the internal temperatures of the pre-stellar cores to generate an updated core mass function (CMF). This updated CMF is consistent with previous results in so far as they went, but our deeper map gives an improved completeness limit of 0.1M$_{\sun}$ (0.16 Jy), which enables us to show that a turnover exists in the low-mass regime of the CMF. The L1688 CMF shows the same form as the stellar IMF and can be mapped onto the stellar IMF, showing that the IMF is determined at the prestellar core stage. We compare L1688 with the Orion star-forming region and find that the turnover in the L1688 CMF occurs at a mass roughly a factor of two lower than the CMF turnover in Orion. This suggests that the position of the CMF turnover may be a function of environment.
\end{abstract}

\section{Introduction}

Star formation in molecular clouds occurs within prestellar cores, which are gravitationally bound cores within the clouds \citep{1994MNRAS.268.276W,1996A&A.314.625A,1999MNRAS.305.143W, 2002Sci...295...76W, 2007prpl.conf...17D, 2007prpl.conf...33W}.

A number of observations have shown that the core mass function (CMF) of prestellar cores appears to mimic \citep[Motte Andre \& Neri 1998, hereafter MAN98,][]{1998ApJ...508L..91T, 2000ApJ...545..327J, 2001A&A...372L..41M, 2001ApJ...559..307J, 2002Sci...295...82K, 2002ApJ...575..950O, 2006ApJ...639..259J} the stellar initial mass function \citep[IMF;][]{1955ApJ...121..161S}.

However, the comparison between the core mass function and stellar IMF has not often been accurately probed at lower masses. It is more difficult to study this part of the mass domain, but recent results have shown that the CMF exhibits a turnover at lower masses in a manner similar to the IMF \citep{2007MNRAS.374.1413N}.

The Ophiuchus star-forming region is located at a distance of 139~pc \citep{2008AN....329...10M} and is a site of low-mass star formation \citep{1983ApJ...274..698W}. The region consists of two main clouds, L1688 and L1689, which both have extended streamers leading out to lengths of around 10 pc \citep{1989ApJ...338..902L}. Specifically, it is the more massive of the two clouds, L1688, which is studied in this paper, and which is generally known as the Ophiuchus main cloud. Very high star formation rates have been measured here, with 14--40\% of the molecular gas being converted into stars \citep{1977AJ.....82..198V}.

The Ophiuchus cloud has been observed in many wavelengths from the visible to the submilimetre \citep[e.g. MAN98, ][]{1983ApJ...269..182M, 1989ApJ...340..823W, 1992ApJ...401..667A, 1992ApJ...395..516G, 1997ApJS..112..109B, 2000ApJ...545..327J, 2001MNRAS.323.1025J, 2004ApJ...611L..45J, 2008ApJS..175..277D}. Because of this, the properties of the cloud are very well known and it is therefore a good place to probe low-mass star formation.

The Ophiuchus cloud is the nearest example of  `clustered' star formation (MAN98). This is important to study because most stars form in clustered environments \citep{1993prpl.conf..429Z}. Ophiuchus may also be the nearest example of triggered star formation in action \citep{1989ApJ...338..902L,2006MNRAS.368.1833N}, making it a prime candidate for study.

\begin{figure*}
\includegraphics[angle=0,width=0.85\textwidth]{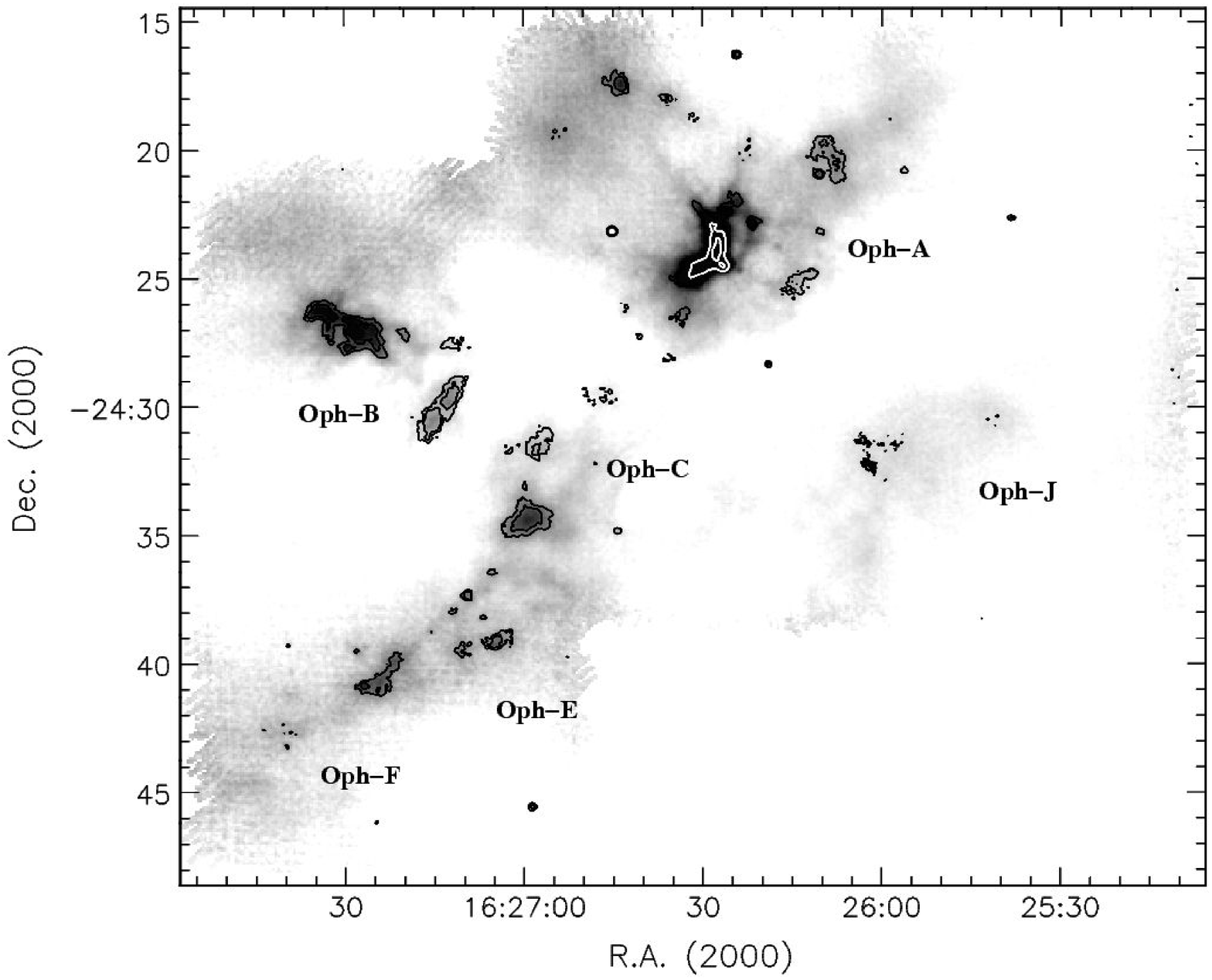}
\caption{Greyscale image and isophotal contour map of the SCUBA 850$\mu$m continuum scan-map data of the Ophiuchus dark cloud L1688. Signal-to-noise contours at 5$\sigma$ and 10$\sigma$ are shown in black; 25$\sigma$ and 100$\sigma$ contours shown in white. 1$\sigma$ noise levels vary from 15 to 40 mJy/beam (see text for details).}
\label{whole-map}
\end{figure*}

\begin{figure}
\includegraphics[angle=0,width=0.40\textwidth]{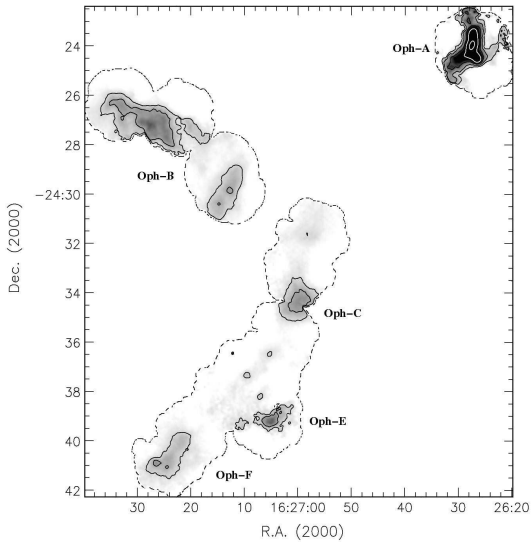}
\caption{Greyscale image and isophotal contour map of the SCUBA 850$\mu$m continuum jiggle map data of the Ophiuchus dark cloud L1688. Contours at 0.25, 0.5 and 1.0 Jy/beam are shown in black; 2.0 and 4.0 Jy/beam contours are shown in white. 1$\sigma$ noise levels vary from 10 to 180 mJy/beam.}
\label{jiggle-map}
\end{figure}

In this study we have combined all of the high signal-to-noise SCUBA wide-field scan-map data and narrow-field jiggle-map data taken of L1688, and re-reduced it to produce the deepest submillimetre map of this cloud ever made. L1688 is the region of the Ophiuchus cloud defined by \citet{1989ApJ...338..902L} and outlined in their Figure~1a (it is marked out by a solid, 5K contour). Of the original regions covered at 1.3mm by MAN98 (Oph-A, -B, -C, -D, -E and -F) only one, Oph-D, is not included here, as it is not part of the central region of L1688. A newly discovered region, which we name Oph-J, is discussed. Two smaller regions, Oph-H and Oph-I are discussed by \citet{2004ApJ...611L..45J} but are not included in this study. We produce a CMF and investigate the low-mass end of the CMF of the cloud. We compare this with previous findings and also with the Orion molecular cloud \citep{2007MNRAS.374.1413N}.

\section{Observations}
\label{obs-sec}

\begin{figure*}
\begin{center}
\subfigure[Region Oph-A.]{
\includegraphics[angle=0,width=68mm]{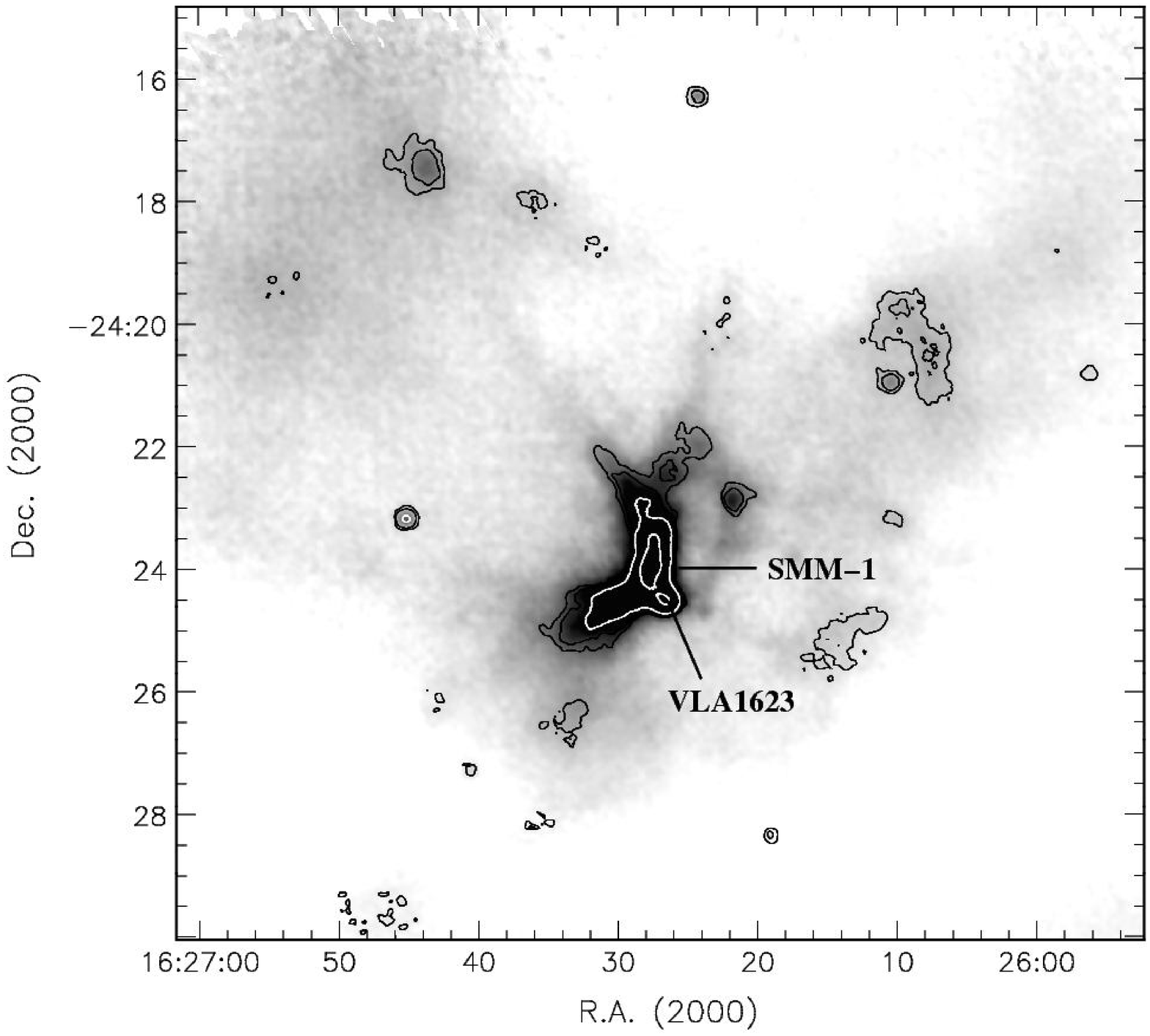}}
\subfigure[Region Oph-B.]{
\includegraphics[angle=0,width=68mm]{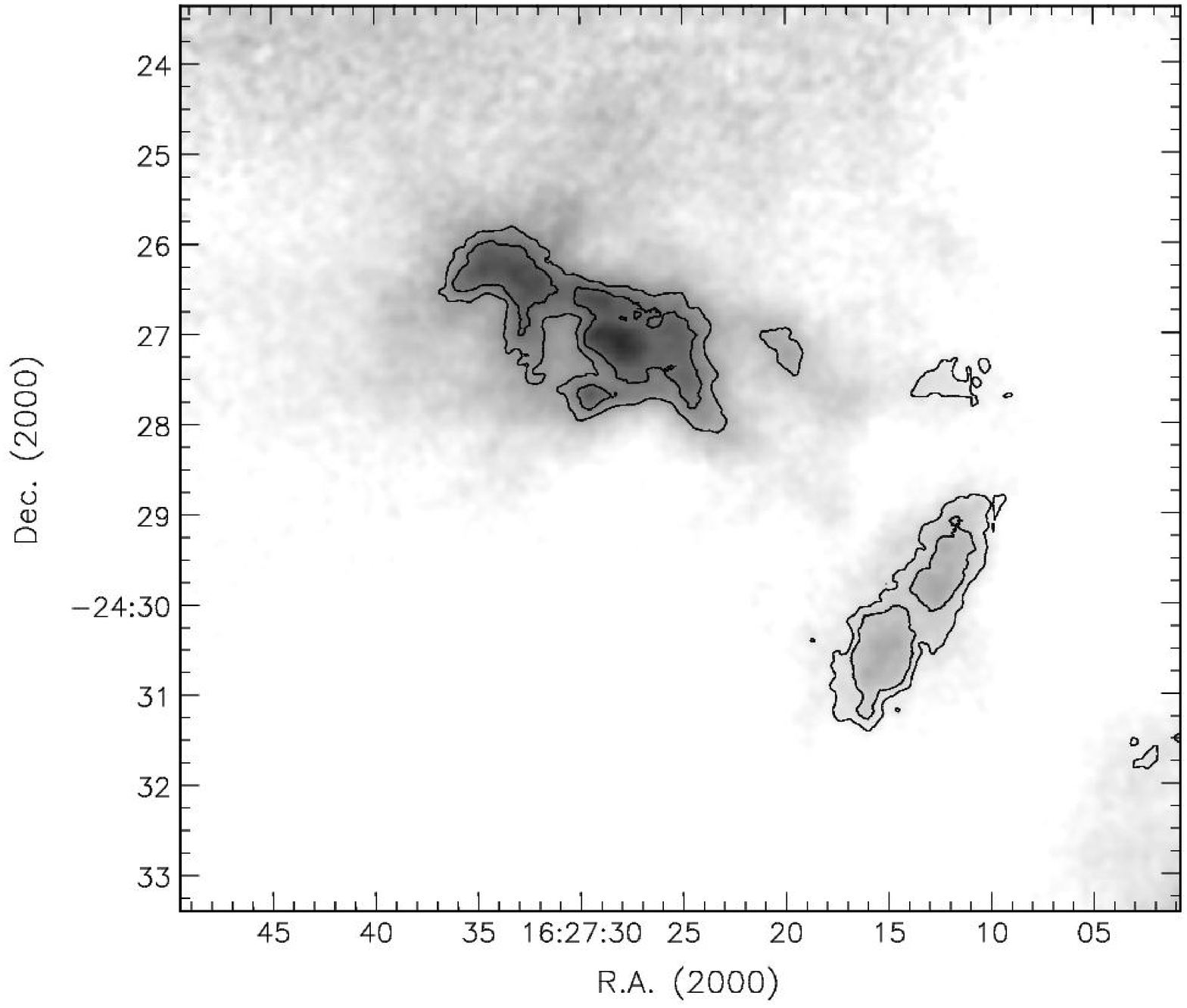}}
\subfigure[Region Oph-C.]{
\includegraphics[angle=0,width=68mm]{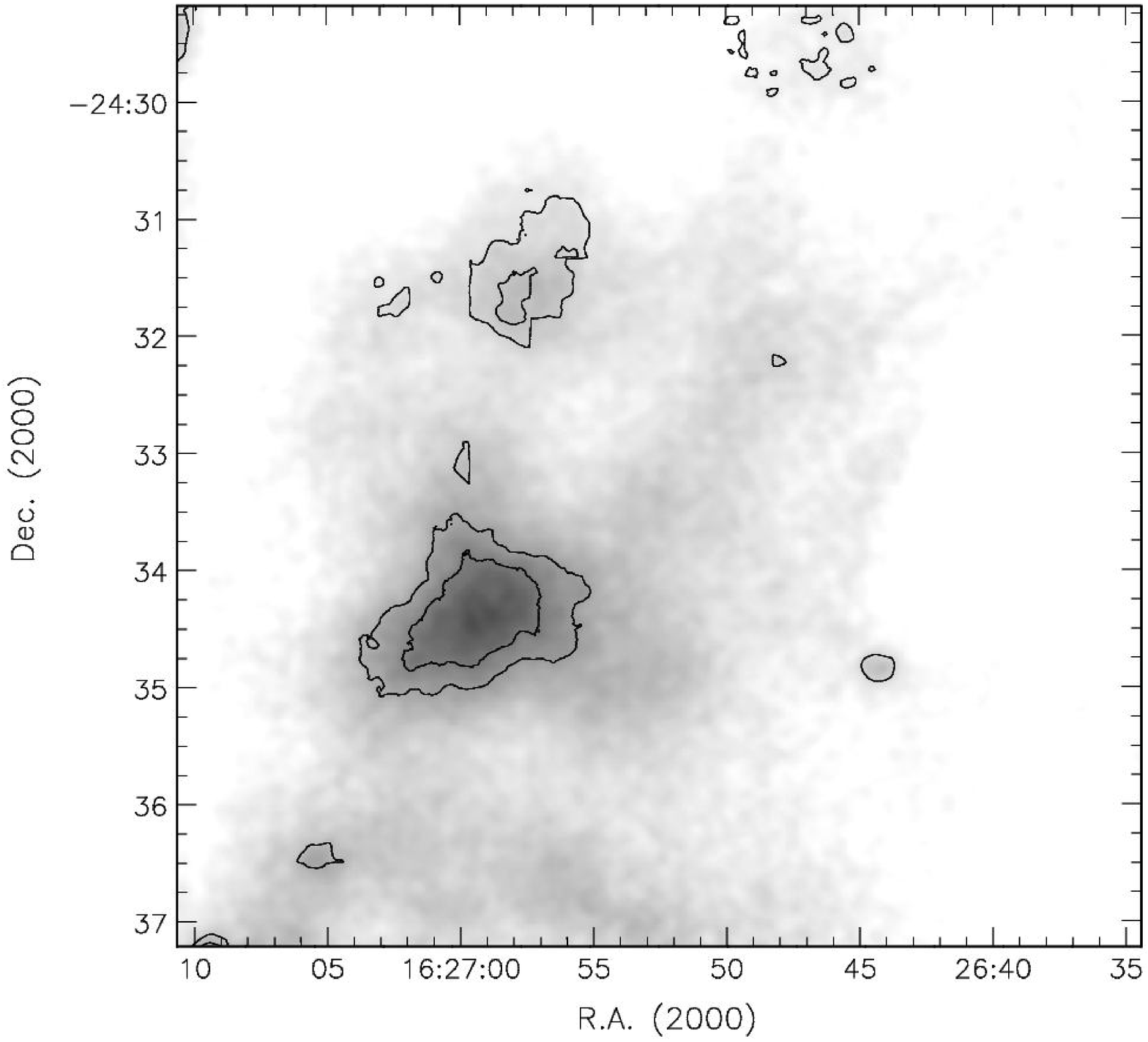}}
\subfigure[Region Oph-E.]{
\includegraphics[angle=0,width=68mm]{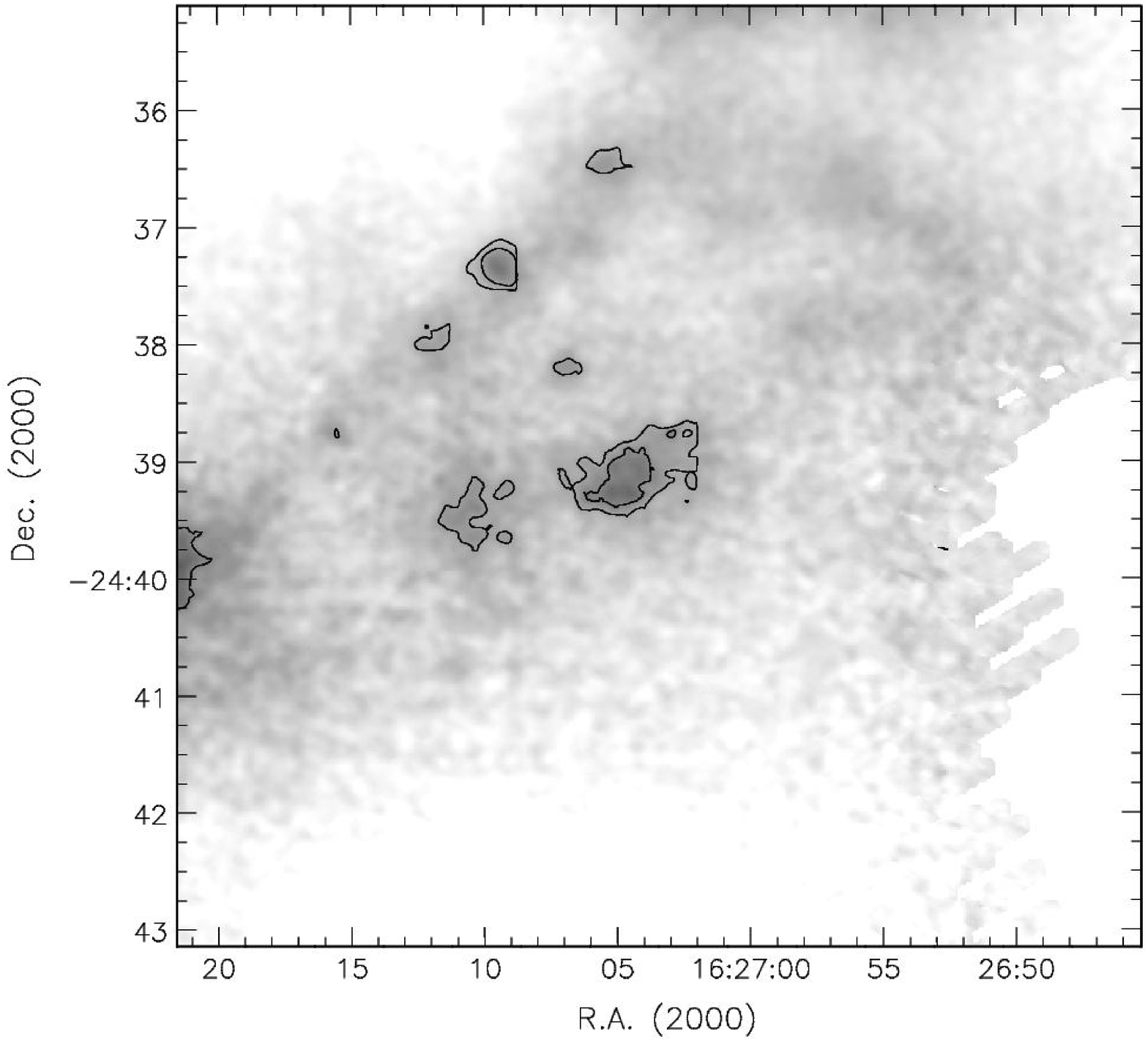}}
\subfigure[Region Oph-F.]{
\includegraphics[angle=0,width=68mm]{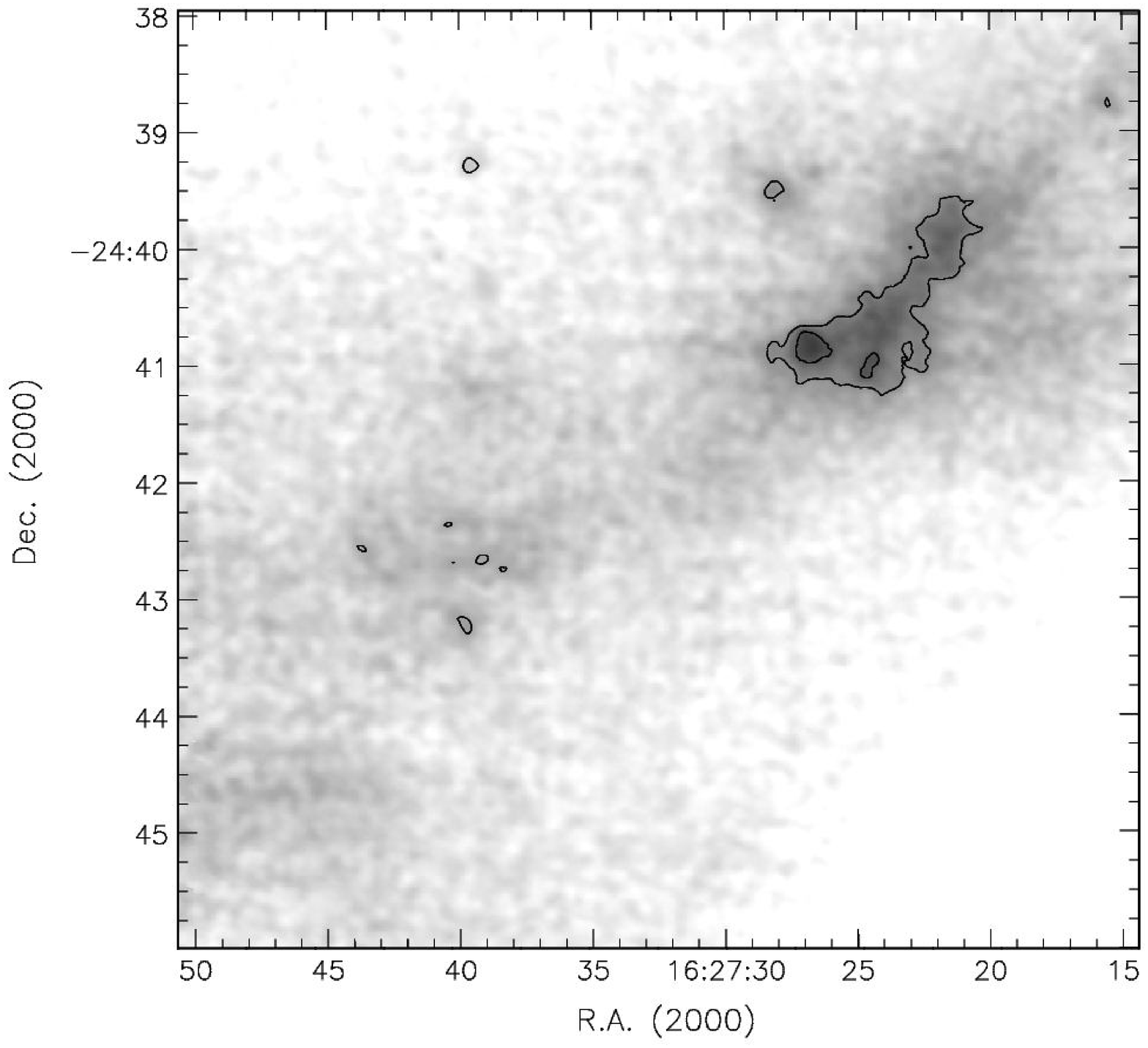}}
\subfigure[Region Oph-J.]{
\includegraphics[angle=0,width=68mm]{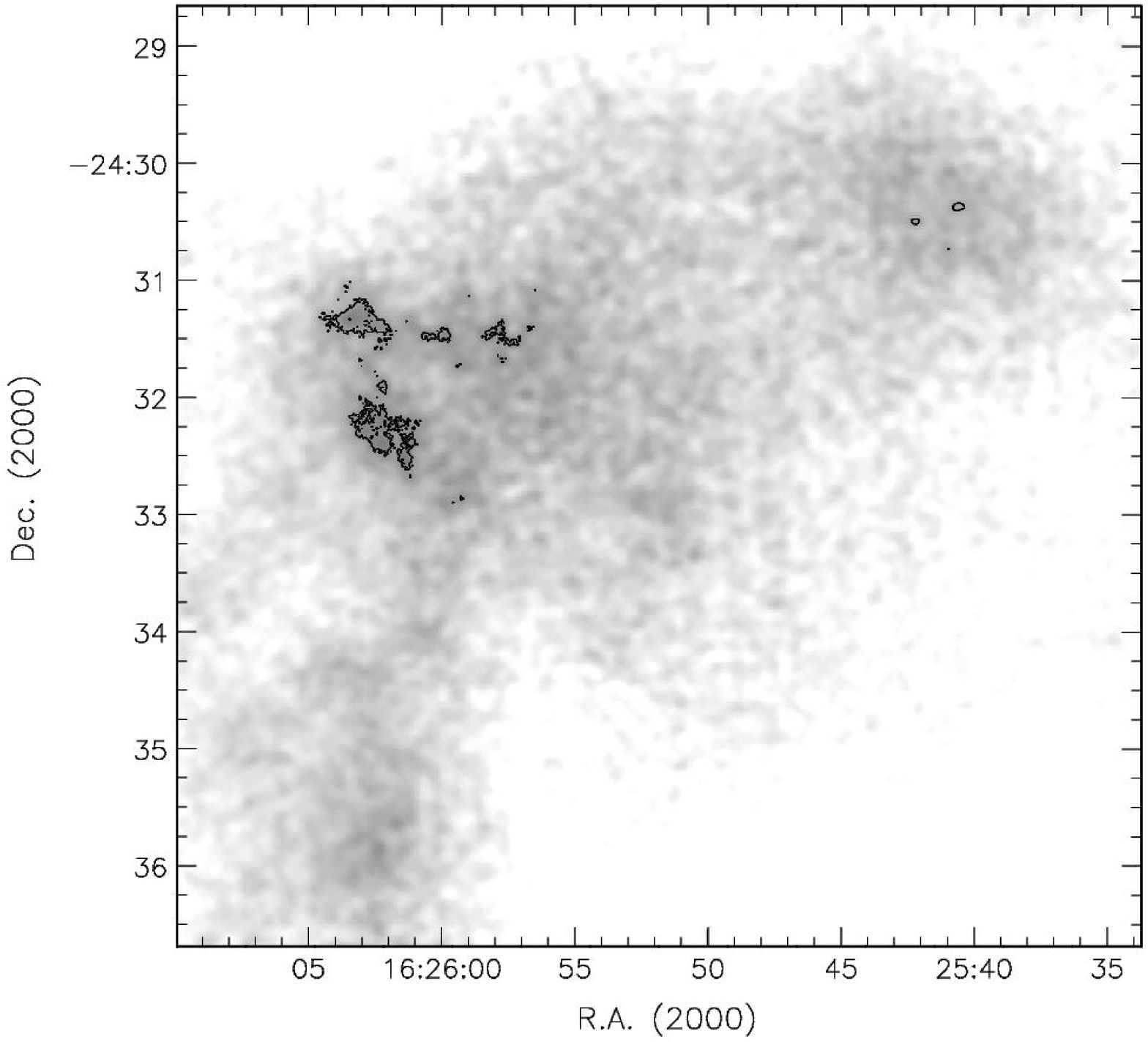}}
\caption{850$\mu$m continuum maps of regions Oph-A, B, C, E, F and new region J. Signal-to-noise contours at 5$\sigma$ and 10$\sigma$ are shown in black; 25$\sigma$ and 100$\sigma$ contours are shown in white.}
\label{region-maps}
\end{center}
\end{figure*}

The submillimetre data presented in this study were obtained using the Submillimetre Common User Bolometer Array (SCUBA) on the James Clerk Maxwell Telescope (JCMT). SCUBA used a dichroic beam-splitter to simultaneously observe at 850~$\mu$m and 450~$\mu$m wavelengths at resolutions of 14 arcsec and 8 arcsec respectively. The data presented here were acquired from the Canadian Astronomy Data Center's JCMT data archive \citep{1997ASPC..125..397T}. Some of these data have been published previously \citep{2000ApJ...545..327J, 2004ApJ...611L..45J, 2008ApJS..175..277D} but have all been consistently re-reduced here using a single method for this study. Some data from these studies has been omitted where the mapping technique is low signal-to-noise. Only the 850~$\mu$m data have been used and are presented here.

SCUBA was used in scan-map mode to produce the observations in Figure~\ref{whole-map}. To acquire a scan-map, the SCUBA array was scanned across the sky at 15.5$^\circ$ to the main axis to achieve Nyquist sampling. It was then able to raster across the sky in this mode and achieve maps several arcminutes across \citep{2000ASPC..217..205J}.

The scan-map data consist of 76 observations over 3 nights at the JCMT on 1998 July 11 and 12, and 1999 March 4 (see Table~\ref{TabObs}). The sky opacity at 850~$\mu$m varied from 0.10 to 0.29 with a mean value of 0.17, as determined by the `skydip' method \citep{2002MNRAS.336....1A}. The data were reduced using the SCUBA User Redution Facility \citep{2000ASPC..216..559J} and calibrated using observations of Uranus or Mars or of the secondary calibrator CRL618 \citep{1994MNRAS.271...75S}.

The scan-map observation mode results in differential maps which must be reconstructed. This is done in fourier space using the Emerson-2 method \citep{1995ASPC...75.....E, 2001techreport.11.2}. Baseline offsets were removed using the SURF command SCAN\_RLB with the `median' method \citep{2000ASPC..216..559J}. The time-series data were checked for each observation and noisy bolometers were removed by eye. We estimate that the calibration uncertainty is $\pm$10\% based on the variation in the calibration factors from Uranus and Mars across all the datasets (see Tables~\ref{TabMass1} and \ref{TabMass2}).

The data make up one map which is split into 7 regions, following the naming methodology in MAN98. Figure~\ref{whole-map} shows the whole map with the individual regions named. \citet{2000ApJ...545..327J} first published the 850~$\mu$m scan-map data used in this study. They found good agreement with the results of MAN98.

SCUBA was used in jiggle-map mode for the observations in Figure~\ref{jiggle-map}. The jiggle-map data consist of 42 observations over 19 nights at the JCMT between 1997 August 7 and 2004 March 29 (see Table~\ref{TabObs}). The sky opacity at 850~$\mu$m varied from 0.21 to 0.52 with a mean value of 0.30, as determined by the `skydip' method. The jiggle-map data were also reduced using the SCUBA User Redution Facility and calibrated using observations of Uranus or Mars or of the secondary calibrator CRL618. These data are shown in  Figure~\ref{jiggle-map}. A close-up of each of the main regions is shown in Figure~\ref{region-maps}.

\begin{table}
\centering {
\caption{Details of observations.}
\label{TabObs}
\begin{tabular}{cccc}
\hline
Date & Mode & $\overline{\tau}$$_{850}$ & No. of\\
 & & & Datasets\\
\hline

1997 Aug 07 & Jiggle Map & 0.285 & 1 \\ 
1997 Aug 08 & Jiggle Map & 0.282 & 1 \\ 
1997 Aug 09 & Jiggle Map & 0.280 & 5 \\ 
1997 Aug 10 & Jiggle Map & 0.220 & 1 \\ 
1997 Aug 11 & Jiggle Map & 0.103 & 1 \\ 
1997 Aug 25 & Jiggle Map & 0.206 & 2 \\ 
1997 Aug 26 & Jiggle Map & 0.183 & 2 \\ 
1997 Sep 09 & Jiggle Map & 0.108 & 6 \\ 
1998 Jun 26 & Jiggle Map & 0.278 & 2 \\ 
1998 Jul 11 & Scan Map & 0.103 & 27 \\ 
1998 Jul 12 & Scan Map & 0.148 & 24 \\ 
1998 Aug 11 & Jiggle Map & 0.211 & 2 \\ 
1998 Aug 25 & Jiggle Map & 0.314 & 2 \\ 
1999 Mar 04 & Scan Map & 0.205 & 25 \\ 
1999 Mar 13 & Jiggle Map & 0.208 & 2 \\ 
1999 Aug 20 & Jiggle Map & 0.279 & 2 \\ 
2000 Aug 08 & Jiggle Map & 0.208 & 1 \\ 
2001 Jul 31 & Jiggle Map & 0.310 & 3 \\ 
2003 Feb 04 & Jiggle Map & 0.260 & 8 \\ 
2004 Mar 29 & Jiggle Map & 0.255 & 1 \\ 
\hline

\end{tabular}
}
\end{table}

\begin{table*}
\caption{Source names, peak positions, 850-$\mu$m peak flux densities
and signal-to-noise ratios of the cores in regions Oph-A, -B, -C, -E, -F, -J. The flux densities S$_{1.3mm}^{peak}$ and S$_{850}^{peak}$ refer to the MAN98 peak flux and the peak flux in our  dataset respectively.}
\label{Tab1}
\begin{tabular}{lcccccrcc}
\hline
Source & \multicolumn{2}{c}{MAN98 Position} & \multicolumn{2}{c}{SCUBA Position}
& \multicolumn{2}{c}{Flux Densities} & \\
Name &  RA    &  Dec. &  RA & Dec  &
$S_{1.3mm}^{peak}$  &  $S_{850}^{peak}$~~~~ & S/N & IR \\
  &  (2000)     &   (2000)    &    (2000)  & (2000) & 
    (mJy/beam)          &     (mJy/beam)         & ($\sigma$) 
  &  assn  \\ \hline

A-MM1 & 16$^{\rm h}$26$^{\rm m}$22.44$^{\rm s}$ & $-$24$^\circ$23$^\prime$40.26$^{\prime\prime}$ & 16$^{\rm h}$26$^{\rm m}$23.00$^{\rm s}$ & $-$24$^\circ$23$^\prime$34.46$^{\prime\prime}$ & 50 $\pm$ 15 & 380 $\pm$ 15~~ & 23 & \\
A-MM2/3 & --- & --- & 16$^{\rm h}$26$^{\rm m}$23.80$^{\rm s}$ & $-$24$^\circ$24$^\prime$09.95$^{\prime\prime}$ &  ---  & 660 $\pm$ 25~~ & 24 & \\
A-MM4 & 16$^{\rm h}$26$^{\rm m}$24.40$^{\rm s}$ & $-$24$^\circ$21$^\prime$52.13$^{\prime\prime}$ & 16$^{\rm h}$26$^{\rm m}$24.14$^{\rm s}$ & $-$24$^\circ$21$^\prime$59.30$^{\prime\prime}$ & 80 $\pm$ 15 & 930 $\pm$ 20~~ & 42 & \\
A-MM5 & 16$^{\rm h}$26$^{\rm m}$26.42$^{\rm s}$ & $-$24$^\circ$22$^\prime$27.00$^{\prime\prime}$ & 16$^{\rm h}$26$^{\rm m}$26.66$^{\rm s}$ & $-$24$^\circ$22$^\prime$28.60$^{\prime\prime}$ & 100 $\pm$ 20 & 1590 $\pm$ 20~~ & 71 & \\
A-MM6 & 16$^{\rm h}$26$^{\rm m}$28.43$^{\rm s}$ & $-$24$^\circ$22$^\prime$52.86$^{\prime\prime}$ & 16$^{\rm h}$26$^{\rm m}$28.21$^{\rm s}$ & $-$24$^\circ$23$^\prime$00.10$^{\prime\prime}$ & 200 $\pm$ 25 & 4050 $\pm$ 30~~ & 138 & \\
A-MM7 & 16$^{\rm h}$26$^{\rm m}$29.42$^{\rm s}$ & $-$24$^\circ$22$^\prime$33.80$^{\prime\prime}$ & 16$^{\rm h}$26$^{\rm m}$29.45$^{\rm s}$ & $-$24$^\circ$22$^\prime$33.01$^{\prime\prime}$ & 110 $\pm$ 30 & 1770 $\pm$ 30~~ & 60 & \\
A-MM8 & 16$^{\rm h}$26$^{\rm m}$33.48$^{\rm s}$ & $-$24$^\circ$25$^\prime$00.53$^{\prime\prime}$ & 16$^{\rm h}$26$^{\rm m}$31.80$^{\rm s}$ & $-$24$^\circ$24$^\prime$50.00$^{\prime\prime}$ & 80 $\pm$ 25 & 6110 $\pm$ 25~~ & 245 & \\
A-MM9 & --- & --- & 16$^{\rm h}$26$^{\rm m}$45.18$^{\rm s}$ & $-$24$^\circ$23$^\prime$10.02$^{\prime\prime}$ &  ---  & 490 $\pm$ 30~~ & 16 & Y\\
A-MM10 & --- & --- & 16$^{\rm h}$26$^{\rm m}$21.83$^{\rm s}$ & $-$24$^\circ$22$^\prime$51.96$^{\prime\prime}$ &  ---  & 1640 $\pm$ 25~~ & 66 & Y\\
A-MM11 & --- & --- & 16$^{\rm h}$26$^{\rm m}$32.74$^{\rm s}$ & $-$24$^\circ$26$^\prime$14.40$^{\prime\prime}$ &  ---  & 1240 $\pm$ 25~~ & 50 & \\
A-MM12 & --- & --- & 16$^{\rm h}$26$^{\rm m}$14.24$^{\rm s}$ & $-$24$^\circ$25$^\prime$04.33$^{\prime\prime}$ &  ---  & 870 $\pm$ 25~~ & 35 & \\
A-MM15 & --- & --- & 16$^{\rm h}$26$^{\rm m}$40.50$^{\rm s}$ & $-$24$^\circ$27$^\prime$16.28$^{\prime\prime}$ &  ---  & 20 $\pm$ 30~~ & 1 & Y\\
A-MM16 & --- & --- & 16$^{\rm h}$26$^{\rm m}$36.26$^{\rm s}$ & $-$24$^\circ$28$^\prime$12.84$^{\prime\prime}$ &  ---  & 60 $\pm$ 20~~ & 3 & \\
A-MM17 & --- & --- & 16$^{\rm h}$26$^{\rm m}$34.77$^{\rm s}$ & $-$24$^\circ$28$^\prime$08.10$^{\prime\prime}$ &  ---  & 60 $\pm$ 20~~ & 3 & \\
A-MM18 & --- & --- & 16$^{\rm h}$26$^{\rm m}$43.73$^{\rm s}$ & $-$24$^\circ$17$^\prime$25.74$^{\prime\prime}$ &  ---  & 1680 $\pm$ 30~~ & 56 & \\
A-MM19 & --- & --- & 16$^{\rm h}$26$^{\rm m}$24.30$^{\rm s}$ & $-$24$^\circ$16$^\prime$16.11$^{\prime\prime}$ &  ---  & 430 $\pm$ 40~~ & 11 & Y\\
A-MM20 & --- & --- & 16$^{\rm h}$26$^{\rm m}$36.18$^{\rm s}$ & $-$24$^\circ$17$^\prime$56.34$^{\prime\prime}$ &  ---  & 240 $\pm$ 30~~ & 8 & \\
A-MM21 & --- & --- & 16$^{\rm h}$26$^{\rm m}$31.64$^{\rm s}$ & $-$24$^\circ$18$^\prime$38.05$^{\prime\prime}$ &  ---  & 410 $\pm$ 35~~ & 12 & \\
A-MM22 & --- & --- & 16$^{\rm h}$26$^{\rm m}$31.45$^{\rm s}$ & $-$24$^\circ$18$^\prime$52.05$^{\prime\prime}$ &  ---  & 210 $\pm$ 35~~ & 6 & \\
A-MM23 & --- & --- & 16$^{\rm h}$26$^{\rm m}$07.89$^{\rm s}$ & $-$24$^\circ$20$^\prime$30.50$^{\prime\prime}$ &  ---  & 2370 $\pm$ 25~~ & 95 & \\
A-MM24 & --- & --- & 16$^{\rm h}$26$^{\rm m}$10.50$^{\rm s}$ & $-$24$^\circ$20$^\prime$56.83$^{\prime\prime}$ &  ---  & 760 $\pm$ 25~~ & 30 & Y\\
A-MM25 & --- & --- & 16$^{\rm h}$25$^{\rm m}$55.96$^{\rm s}$ & $-$24$^\circ$20$^\prime$49.47$^{\prime\prime}$ &  ---  & 140 $\pm$ 30~~ & 5 & Y\\
A-MM26 & --- & --- & 16$^{\rm h}$26$^{\rm m}$15.40$^{\rm s}$ & $-$24$^\circ$25$^\prime$32.50$^{\prime\prime}$ &  ---  & 820 $\pm$ 25~~ & 33 & \\
A-MM27 & --- & --- & 16$^{\rm h}$26$^{\rm m}$13.85$^{\rm s}$ & $-$24$^\circ$25$^\prime$25.16$^{\prime\prime}$ &  ---  & 650 $\pm$ 25~~ & 26 & \\
A-MM28 & --- & --- & 16$^{\rm h}$26$^{\rm m}$54.75$^{\rm s}$ & $-$24$^\circ$19$^\prime$16.55$^{\prime\prime}$ &  ---  & 410 $\pm$ 40~~ & 10 & \\
A-MM29 & --- & --- & 16$^{\rm h}$26$^{\rm m}$53.36$^{\rm s}$ & $-$24$^\circ$19$^\prime$28.08$^{\prime\prime}$ &  ---  & 350 $\pm$ 40~~ & 9 & \\
A-MM30 & --- & --- & 16$^{\rm h}$26$^{\rm m}$09.63$^{\rm s}$ & $-$24$^\circ$19$^\prime$43.25$^{\prime\prime}$ &  ---  & 2460 $\pm$ 20~~ & 123 & \\
A2-MM1 & 16$^{\rm h}$26$^{\rm m}$11.45$^{\rm s}$ & $-$24$^\circ$24$^\prime$40.00$^{\prime\prime}$ & 16$^{\rm h}$26$^{\rm m}$11.73$^{\rm s}$ & $-$24$^\circ$24$^\prime$54.16$^{\prime\prime}$ & 60 $\pm$ 10 & 640 $\pm$ 20~~ & 32 & \\
A3-MM1 & 16$^{\rm h}$26$^{\rm m}$09.41$^{\rm s}$ & $-$23$^\circ$24$^\prime$06.13$^{\prime\prime}$ & 16$^{\rm h}$26$^{\rm m}$10.07$^{\rm s}$ & $-$24$^\circ$23$^\prime$11.00$^{\prime\prime}$ & 90 $\pm$ 10 & 640 $\pm$ 30~~ & 21 & \\
A-N & 16$^{\rm h}$26$^{\rm m}$21.35$^{\rm s}$ & $-$24$^\circ$19$^\prime$40.33$^{\prime\prime}$ & 16$^{\rm h}$26$^{\rm m}$22.74$^{\rm s}$ & $-$24$^\circ$20$^\prime$00.00$^{\prime\prime}$ & 60 $\pm$ 10 & 410 $\pm$ 30~~ & 14 & \\
A-S & --- & --- & 16$^{\rm h}$26$^{\rm m}$42.69$^{\rm s}$ & $-$24$^\circ$26$^\prime$08.05$^{\prime\prime}$ & 85 $\pm$ 10 & 10 $\pm$ 35~~ & 0 & \\
SM1 & 16$^{\rm h}$26$^{\rm m}$27.45$^{\rm s}$ & $-$24$^\circ$23$^\prime$55.93$^{\prime\prime}$ & 16$^{\rm h}$26$^{\rm m}$27.73$^{\rm s}$ & $-$24$^\circ$23$^\prime$58.17$^{\prime\prime}$ & 1300 $\pm$ 20 & 14140 $\pm$ 25~~ & 571 & \\
SM1N & 16$^{\rm h}$26$^{\rm m}$27.44$^{\rm s}$ & $-$24$^\circ$23$^\prime$27.93$^{\prime\prime}$ & 16$^{\rm h}$26$^{\rm m}$27.46$^{\rm s}$ & $-$24$^\circ$23$^\prime$32.71$^{\prime\prime}$ & 790 $\pm$ 25 & 5600 $\pm$ 15~~ & 359 & \\
SM2 & 16$^{\rm h}$26$^{\rm m}$29.46$^{\rm s}$ & $-$24$^\circ$24$^\prime$25.91$^{\prime\prime}$ & 16$^{\rm h}$26$^{\rm m}$29.41$^{\rm s}$ & $-$24$^\circ$24$^\prime$26.69$^{\prime\prime}$ & 450 $\pm$ 30 & 11490 $\pm$ 15~~ & 690 & \\
VLA1623 & --- & --- & 16$^{\rm h}$26$^{\rm m}$26.74$^{\rm s}$ & $-$24$^\circ$24$^\prime$30.00$^{\prime\prime}$ &  ---  & 5630 $\pm$ 30~~ & 188 & \\

\hline

B1-MM1 & 16$^{\rm h}$27$^{\rm m}$08.57$^{\rm s}$ & $-$24$^\circ$27$^\prime$50.19$^{\prime\prime}$ & 16$^{\rm h}$27$^{\rm m}$09.32$^{\rm s}$ & $-$24$^\circ$27$^\prime$43.73$^{\prime\prime}$ & 50 $\pm$ 5 & 50 $\pm$ 30~~ & 2 & \\
B1-MM2 & 16$^{\rm h}$27$^{\rm m}$11.60$^{\rm s}$ & $-$24$^\circ$29$^\prime$17.99$^{\prime\prime}$ & 16$^{\rm h}$27$^{\rm m}$12.23$^{\rm s}$ & $-$24$^\circ$29$^\prime$23.65$^{\prime\prime}$ & 45 $\pm$ 10 & 2140 $\pm$ 10~~ & 230 & \\
B1-MM3 & 16$^{\rm h}$27$^{\rm m}$12.62$^{\rm s}$ & $-$24$^\circ$29$^\prime$57.92$^{\prime\prime}$ & 16$^{\rm h}$27$^{\rm m}$12.60$^{\rm s}$ & $-$24$^\circ$29$^\prime$49.89$^{\prime\prime}$ & 65 $\pm$ 10 & 4140 $\pm$ 10~~ & 446 & \\
B1-MM4 & 16$^{\rm h}$27$^{\rm m}$15.64$^{\rm s}$ & $-$24$^\circ$30$^\prime$41.72$^{\prime\prime}$ & 16$^{\rm h}$27$^{\rm m}$15.32$^{\rm s}$ & $-$24$^\circ$30$^\prime$36.82$^{\prime\prime}$ & 60 $\pm$ 15 & 3330 $\pm$ 10~~ & 372 & \\
B1-MM5 & --- & --- & 16$^{\rm h}$27$^{\rm m}$16.05$^{\rm s}$ & $-$24$^\circ$31$^\prime$09.30$^{\prime\prime}$ &  ---  & 650 $\pm$ 25~~ & 26 & Y\\
B1-MM6 & --- & --- & 16$^{\rm h}$27$^{\rm m}$10.58$^{\rm s}$ & $-$24$^\circ$28$^\prime$54.69$^{\prime\prime}$ &  ---  & 350 $\pm$ 30~~ & 12 & \\
B1-MM7 & --- & --- & 16$^{\rm h}$27$^{\rm m}$18.72$^{\rm s}$ & $-$24$^\circ$30$^\prime$24.64$^{\prime\prime}$ &  ---  & 180 $\pm$ 30~~ & 6 & \\
B1B2-MM1 & 16$^{\rm h}$27$^{\rm m}$11.57$^{\rm s}$ & $-$24$^\circ$27$^\prime$38.99$^{\prime\prime}$ & 16$^{\rm h}$27$^{\rm m}$12.44$^{\rm s}$ & $-$24$^\circ$27$^\prime$30.31$^{\prime\prime}$ & 40 $\pm$ 5 & 350 $\pm$ 35~~ & 10 & \\
B1B2-MM2 & 16$^{\rm h}$27$^{\rm m}$17.60$^{\rm s}$ & $-$24$^\circ$28$^\prime$47.59$^{\prime\prime}$ & 16$^{\rm h}$27$^{\rm m}$17.77$^{\rm s}$ & $-$24$^\circ$28$^\prime$59.75$^{\prime\prime}$ & 45 $\pm$ 10 & 490 $\pm$ 10~~ & 51 & Y\\
B2-MM2 & 16$^{\rm h}$27$^{\rm m}$20.56$^{\rm s}$ & $-$24$^\circ$27$^\prime$08.39$^{\prime\prime}$ & 16$^{\rm h}$27$^{\rm m}$19.82$^{\rm s}$ & $-$24$^\circ$27$^\prime$13.87$^{\prime\prime}$ & 85 $\pm$ 10 & 1390 $\pm$ 20~~ & 79 & \\
B2-MM4 & 16$^{\rm h}$27$^{\rm m}$24.58$^{\rm s}$ & $-$24$^\circ$27$^\prime$45.12$^{\prime\prime}$ & 16$^{\rm h}$27$^{\rm m}$24.50$^{\rm s}$ & $-$24$^\circ$27$^\prime$46.30$^{\prime\prime}$ & 90 $\pm$ 15 & 1520 $\pm$ 20~~ & 86 & \\
B2-MM5 & 16$^{\rm h}$27$^{\rm m}$24.57$^{\rm s}$ & $-$24$^\circ$27$^\prime$26.12$^{\prime\prime}$ & 16$^{\rm h}$27$^{\rm m}$24.74$^{\rm s}$ & $-$24$^\circ$27$^\prime$29.29$^{\prime\prime}$ & 100 $\pm$ 15 & 1810 $\pm$ 20~~ & 102 & \\
B2-MM6 & 16$^{\rm h}$27$^{\rm m}$25.57$^{\rm s}$ & $-$24$^\circ$27$^\prime$00.05$^{\prime\prime}$ & 16$^{\rm h}$27$^{\rm m}$25.57$^{\rm s}$ & $-$24$^\circ$26$^\prime$58.19$^{\prime\prime}$ & 150 $\pm$ 15 & 2570 $\pm$ 20~~ & 134 & \\
B2-MM7 & 16$^{\rm h}$27$^{\rm m}$27.58$^{\rm s}$ & $-$24$^\circ$27$^\prime$38.92$^{\prime\prime}$ & 16$^{\rm h}$27$^{\rm m}$27.70$^{\rm s}$ & $-$24$^\circ$27$^\prime$38.86$^{\prime\prime}$ & 100 $\pm$ 20 & 990 $\pm$ 25~~ & 39 & \\
B2-MM8 & 16$^{\rm h}$27$^{\rm m}$27.57$^{\rm s}$ & $-$24$^\circ$27$^\prime$06.92$^{\prime\prime}$ & 16$^{\rm h}$27$^{\rm m}$27.96$^{\rm s}$ & $-$24$^\circ$27$^\prime$06.85$^{\prime\prime}$ & 215 $\pm$ 20 & 3140 $\pm$ 20~~ & 178 & \\
B2-MM9 & 16$^{\rm h}$27$^{\rm m}$28.56$^{\rm s}$ & $-$24$^\circ$26$^\prime$36.85$^{\prime\prime}$ & 16$^{\rm h}$27$^{\rm m}$28.82$^{\rm s}$ & $-$24$^\circ$26$^\prime$38.59$^{\prime\prime}$ & 110 $\pm$ 15 & 2170 $\pm$ 20~~ & 113 & \\
B2-MM10 & 16$^{\rm h}$27$^{\rm m}$29.58$^{\rm s}$ & $-$24$^\circ$27$^\prime$41.78$^{\prime\prime}$ & 16$^{\rm h}$27$^{\rm m}$29.53$^{\rm s}$ & $-$24$^\circ$27$^\prime$40.85$^{\prime\prime}$ & 160 $\pm$ 10 & 1020 $\pm$ 25~~ & 41 & Y\\
B2-MM13 & 16$^{\rm h}$27$^{\rm m}$32.55$^{\rm s}$ & $-$24$^\circ$26$^\prime$06.58$^{\prime\prime}$ & 16$^{\rm h}$27$^{\rm m}$32.95$^{\rm s}$ & $-$24$^\circ$26$^\prime$03.16$^{\prime\prime}$ & 75 $\pm$ 15 & 780 $\pm$ 30~~ & 26 & \\
B2-MM14 & 16$^{\rm h}$27$^{\rm m}$32.56$^{\rm s}$ & $-$24$^\circ$26$^\prime$28.58$^{\prime\prime}$ & 16$^{\rm h}$27$^{\rm m}$32.58$^{\rm s}$ & $-$24$^\circ$26$^\prime$27.46$^{\prime\prime}$ & 130 $\pm$ 15 & 1990 $\pm$ 15~~ & 128 & \\
B2-MM15 & 16$^{\rm h}$27$^{\rm m}$32.57$^{\rm s}$ & $-$24$^\circ$27$^\prime$02.58$^{\prime\prime}$ & 16$^{\rm h}$27$^{\rm m}$32.87$^{\rm s}$ & $-$24$^\circ$26$^\prime$59.16$^{\prime\prime}$ & 90 $\pm$ 15 & 990 $\pm$ 30~~ & 33 & \\
B2-MM16 & 16$^{\rm h}$27$^{\rm m}$34.56$^{\rm s}$ & $-$24$^\circ$26$^\prime$12.45$^{\prime\prime}$ & 16$^{\rm h}$27$^{\rm m}$34.62$^{\rm s}$ & $-$24$^\circ$26$^\prime$16.39$^{\prime\prime}$ & 100 $\pm$ 15 & 2090 $\pm$ 15~~ & 134 & \\

\hline
\end{tabular}
\end{table*}

\begin{table*}
\contcaption{Core properties.}
\label{Tab2}
\begin{tabular}{lcccccrcc}
\hline
Source & \multicolumn{2}{c}{MAN98 Position} & \multicolumn{2}{c}{SCUBA Position}
& \multicolumn{2}{c}{Flux Densities} & \\
Name &  RA    &  Dec. &  RA & Dec  &
$S_{1.3mm}^{peak}$  &  $S_{850}^{peak}$~~~~ & S/N & IR \\
  &  (2000)     &   (2000)    &    (2000)  & (2000) & 
    (mJy/beam)          &     (mJy/beam)         & ($\sigma$) 
  &  assn  \\ \hline

C-MM2 & 16$^{\rm h}$26$^{\rm m}$58.69$^{\rm s}$ & $-$24$^\circ$33$^\prime$52.85$^{\prime\prime}$ & 16$^{\rm h}$26$^{\rm m}$59.75$^{\rm s}$ & $-$24$^\circ$33$^\prime$56.95$^{\prime\prime}$ & 45 $\pm$ 10 & 1760 $\pm$ 25~~ & 70 & \\
C-MM3 & 16$^{\rm h}$26$^{\rm m}$58.70$^{\rm s}$ & $-$24$^\circ$34$^\prime$21.85$^{\prime\prime}$ & 16$^{\rm h}$26$^{\rm m}$58.80$^{\rm s}$ & $-$24$^\circ$34$^\prime$23.40$^{\prime\prime}$ & 55 $\pm$ 25 & 2260 $\pm$ 20~~ & 106 & \\
C-MM5 & 16$^{\rm h}$26$^{\rm m}$59.70$^{\rm s}$ & $-$24$^\circ$34$^\prime$26.79$^{\prime\prime}$ & 16$^{\rm h}$26$^{\rm m}$59.61$^{\rm s}$ & $-$24$^\circ$34$^\prime$26.81$^{\prime\prime}$ & 50 $\pm$ 25 & 1970 $\pm$ 20~~ & 93 & \\
C-MM6 & 16$^{\rm h}$27$^{\rm m}$01.71$^{\rm s}$ & $-$24$^\circ$34$^\prime$36.65$^{\prime\prime}$ & 16$^{\rm h}$27$^{\rm m}$01.58$^{\rm s}$ & $-$24$^\circ$34$^\prime$44.62$^{\prime\prime}$ & 60 $\pm$ 20 & 1670 $\pm$ 20~~ & 79 & \\
C-MM8 & --- & --- & 16$^{\rm h}$26$^{\rm m}$49.09$^{\rm s}$ & $-$24$^\circ$29$^\prime$45.15$^{\prime\prime}$ &  ---  & 240 $\pm$ 35~~ & 7 & \\
C-MM9 & --- & --- & 16$^{\rm h}$26$^{\rm m}$48.10$^{\rm s}$ & $-$24$^\circ$32$^\prime$12.50$^{\prime\prime}$ &  ---  & 90 $\pm$ 30~~ & 3 & \\
C-MM10 & --- & --- & 16$^{\rm h}$27$^{\rm m}$02.26$^{\rm s}$ & $-$24$^\circ$31$^\prime$42.56$^{\prime\prime}$ &  ---  & 580 $\pm$ 25~~ & 23 & \\
C-MM11 & --- & --- & 16$^{\rm h}$26$^{\rm m}$44.27$^{\rm s}$ & $-$24$^\circ$34$^\prime$50.43$^{\prime\prime}$ &  ---  & 110 $\pm$ 30~~ & 4 & Y\\
C-MM12 & --- & --- & 16$^{\rm h}$26$^{\rm m}$59.80$^{\rm s}$ & $-$24$^\circ$33$^\prime$08.75$^{\prime\prime}$ &  ---  & 150 $\pm$ 30~~ & 5 & \\
C-N & 16$^{\rm h}$26$^{\rm m}$57.64$^{\rm s}$ & $-$24$^\circ$31$^\prime$38.92$^{\prime\prime}$ & 16$^{\rm h}$26$^{\rm m}$58.11$^{\rm s}$ & $-$24$^\circ$31$^\prime$46.32$^{\prime\prime}$ & 60 $\pm$ 10 & 2540 $\pm$ 25~~ & 105 & \\

\hline

E-MM2a & 16$^{\rm h}$27$^{\rm m}$01.59$^{\rm s}$ & $-$24$^\circ$38$^\prime$28.66$^{\prime\prime}$ & 16$^{\rm h}$27$^{\rm m}$02.89$^{\rm s}$ & $-$24$^\circ$38$^\prime$46.47$^{\prime\prime}$ & 50 $\pm$ 15 & 200 $\pm$ 40~~ & 5 & \\
E-MM2b & 16$^{\rm h}$27$^{\rm m}$01.80$^{\rm s}$ & $-$24$^\circ$38$^\prime$51.65$^{\prime\prime}$ & 16$^{\rm h}$27$^{\rm m}$02.19$^{\rm s}$ & $-$24$^\circ$39$^\prime$11.49$^{\prime\prime}$ & 60 $\pm$ 15 & 250 $\pm$ 40~~ & 6 & \\
E-MM2d & 16$^{\rm h}$27$^{\rm m}$04.81$^{\rm s}$ & $-$24$^\circ$39$^\prime$15.45$^{\prime\prime}$ & 16$^{\rm h}$27$^{\rm m}$04.64$^{\rm s}$ & $-$24$^\circ$39$^\prime$15.63$^{\prime\prime}$ & 110 $\pm$ 20 & 660 $\pm$ 40~~ & 17 & \\
E-MM4 & 16$^{\rm h}$27$^{\rm m}$10.82$^{\rm s}$ & $-$24$^\circ$39$^\prime$30.05$^{\prime\prime}$ & 16$^{\rm h}$27$^{\rm m}$10.67$^{\rm s}$ & $-$24$^\circ$39$^\prime$25.28$^{\prime\prime}$ & 50 $\pm$ 10 & 490 $\pm$ 40~~ & 13 & \\
E-MM5 & 16$^{\rm h}$27$^{\rm m}$11.79$^{\rm s}$ & $-$24$^\circ$37$^\prime$56.98$^{\prime\prime}$ & 16$^{\rm h}$27$^{\rm m}$11.67$^{\rm s}$ & $-$24$^\circ$37$^\prime$57.46$^{\prime\prime}$ & 55 $\pm$ 10 & 440 $\pm$ 30~~ & 14 & \\
E-MM6 & --- & --- & 16$^{\rm h}$27$^{\rm m}$09.46$^{\rm s}$ & $-$24$^\circ$37$^\prime$20.92$^{\prime\prime}$ &  ---  & 310 $\pm$ 35~~ & 9 & Y\\
E-MM7 & --- & --- & 16$^{\rm h}$27$^{\rm m}$05.45$^{\rm s}$ & $-$24$^\circ$36$^\prime$27.02$^{\prime\prime}$ &  ---  & 200 $\pm$ 30~~ & 7 & Y\\
E-MM8 & --- & --- & 16$^{\rm h}$27$^{\rm m}$04.19$^{\rm s}$ & $-$24$^\circ$39$^\prime$02.59$^{\prime\prime}$ &  ---  & 300 $\pm$ 30~~ & 10 & \\
E-MM9 & --- & --- & 16$^{\rm h}$27$^{\rm m}$06.84$^{\rm s}$ & $-$24$^\circ$38$^\prime$11.38$^{\prime\prime}$ &  ---  & 140 $\pm$ 30~~ & 5 & Y\\
E-MM10 & --- & --- & 16$^{\rm h}$27$^{\rm m}$15.55$^{\rm s}$ & $-$24$^\circ$38$^\prime$45.95$^{\prime\prime}$ &  ---  & 100 $\pm$ 40~~ & 3 & Y\\

\hline

F-MM1 & 16$^{\rm h}$27$^{\rm m}$21.84$^{\rm s}$ & $-$24$^\circ$39$^\prime$59.31$^{\prime\prime}$ & 16$^{\rm h}$27$^{\rm m}$21.49$^{\rm s}$ & $-$24$^\circ$39$^\prime$54.38$^{\prime\prime}$ & 65 $\pm$ 20 & 1660 $\pm$ 20~~ & 74 & \\
F-MM2a & 16$^{\rm h}$27$^{\rm m}$23.86$^{\rm s}$ & $-$24$^\circ$40$^\prime$35.18$^{\prime\prime}$ & 16$^{\rm h}$27$^{\rm m}$24.43$^{\rm s}$ & $-$24$^\circ$40$^\prime$34.79$^{\prime\prime}$ &  ---  & 1760 $\pm$ 25~~ & 69 & \\
F-MM2b & 16$^{\rm h}$27$^{\rm m}$23.86$^{\rm s}$ & $-$24$^\circ$40$^\prime$35.18$^{\prime\prime}$ & 16$^{\rm h}$27$^{\rm m}$24.69$^{\rm s}$ & $-$24$^\circ$41$^\prime$04.03$^{\prime\prime}$ &  ---  & 1220 $\pm$ 25~~ & 48 & Y\\
F-MM3 & --- & --- & 16$^{\rm h}$27$^{\rm m}$26.69$^{\rm s}$ & $-$24$^\circ$40$^\prime$51.71$^{\prime\prime}$ &  ---  & 1080 $\pm$ 35~~ & 31 & Y\\
F-MM4 & --- & --- & 16$^{\rm h}$27$^{\rm m}$28.20$^{\rm s}$ & $-$24$^\circ$39$^\prime$30.17$^{\prime\prime}$ &  ---  & 310 $\pm$ 35~~ & 9 & Y\\
F-MM5 & --- & --- & 16$^{\rm h}$27$^{\rm m}$39.62$^{\rm s}$ & $-$24$^\circ$39$^\prime$15.26$^{\prime\prime}$ &  ---  & 130 $\pm$ 35~~ & 4 & Y\\
F-MM6 & --- & --- & 16$^{\rm h}$27$^{\rm m}$43.73$^{\rm s}$ & $-$24$^\circ$42$^\prime$34.59$^{\prime\prime}$ &  ---  & 150 $\pm$ 40~~ & 4 & \\
F-MM7 & --- & --- & 16$^{\rm h}$27$^{\rm m}$39.81$^{\rm s}$ & $-$24$^\circ$43$^\prime$12.25$^{\prime\prime}$ &  ---  & 190 $\pm$ 40~~ & 5 & Y\\
F-MM8 & --- & --- & 16$^{\rm h}$27$^{\rm m}$39.22$^{\rm s}$ & $-$24$^\circ$42$^\prime$39.77$^{\prime\prime}$ &  ---  & 330 $\pm$ 40~~ & 8 & \\
F-MM9 & --- & --- & 16$^{\rm h}$27$^{\rm m}$40.46$^{\rm s}$ & $-$24$^\circ$42$^\prime$20.22$^{\prime\prime}$ &  ---  & 150 $\pm$ 40~~ & 4 & \\

\hline

J-MM1 & --- & --- & 16$^{\rm h}$26$^{\rm m}$19.12$^{\rm s}$ & $-$24$^\circ$28$^\prime$20.14$^{\prime\prime}$ &  ---  & 60 $\pm$ 15~~ & 4 & Y\\
J-MM2 & --- & --- & 16$^{\rm h}$26$^{\rm m}$02.17$^{\rm s}$ & $-$24$^\circ$32$^\prime$25.49$^{\prime\prime}$ &  ---  & 670 $\pm$ 25~~ & 27 & \\
J-MM3 & --- & --- & 16$^{\rm h}$26$^{\rm m}$03.45$^{\rm s}$ & $-$24$^\circ$31$^\prime$20.50$^{\prime\prime}$ &  ---  & 550 $\pm$ 25~~ & 22 & \\
J-MM4 & --- & --- & 16$^{\rm h}$25$^{\rm m}$59.95$^{\rm s}$ & $-$24$^\circ$31$^\prime$29.82$^{\prime\prime}$ &  ---  & 280 $\pm$ 20~~ & 14 & \\
J-MM5 & --- & --- & 16$^{\rm h}$25$^{\rm m}$40.64$^{\rm s}$ & $-$24$^\circ$30$^\prime$22.68$^{\prime\prime}$ &  ---  & 250 $\pm$ 20~~ & 13 & \\
J-MM6 & --- & --- & 16$^{\rm h}$25$^{\rm m}$42.25$^{\rm s}$ & $-$24$^\circ$30$^\prime$29.70$^{\prime\prime}$ &  ---  & 120 $\pm$ 25~~ & 5 & \\
J-MM7 & --- & --- & 16$^{\rm h}$25$^{\rm m}$38.39$^{\rm s}$ & $-$24$^\circ$22$^\prime$37.82$^{\prime\prime}$ &  ---  & 100 $\pm$ 35~~ & 3 & Y\\

\hline
\end{tabular}
\end{table*}

\section{Results}
\label{SecResults}

The final map was created from data taken over a range of dates and weather conditions at the JCMT. To produce a final map, with quantified noise levels, we use the technique from \citet{2007MNRAS.374.1413N}. The fluxes quoted in Table \ref{Tab1} are the result of combining the scan- and jiggle-map data. The fluxes from coincident scan- and jiggle-map sources were combined using the noise in each map as a weighting factor.

In this technique the bright sources are masked before smoothing the map to remove the large-scale structure. This is then subtracted from the original data. The resultant map is then used to create a noise map by measuring the standard deviation of the pixel values in a series of 50 arcsec apertures. The gaps from the bright sources are filled in by simple interpolation between the edge values.

\begin{figure}
\includegraphics[angle=0,width=0.48\textwidth]{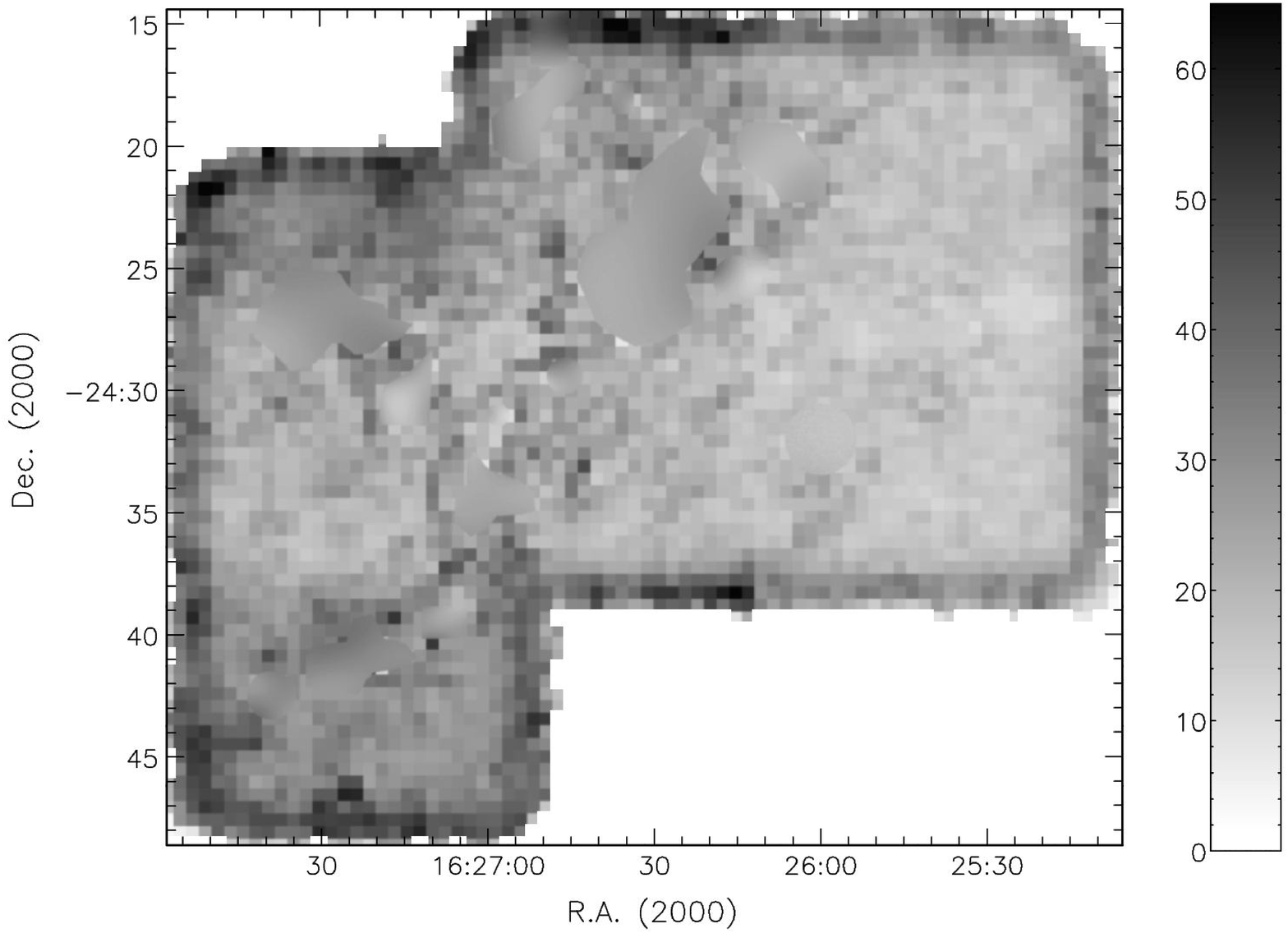}
\caption{The 850$\mu$m noise-map for the Ophiuchus dark cloud L1688. The scale bar shown is in units of mJy/beam.}
\label{noise-map}
\end{figure}

The resultant noise map is shown in Figure~\ref{noise-map}. It can be seen that the south and north-east corners of the map have a higher level of noise. These are the two regions containing the least number of observations. The noise in the whole map ranges from 15 to 40 mJy beam$^{-1}$ with a median value of 25 mJy beam$^{-1}$.

A signal-to-noise map was created by dividing the data map by the noise map. Sources were identified with the method from \citet{2007MNRAS.374.1413N}, in which the signal-to-noise map is used along with the following set of criteria. Any sources with a peak flux density of 5$\sigma$ or more were taken to be real. A dip of at least 3$\sigma$ was required between two adjacent peaks for those two peaks to be identified as two separate sources.

\begin{figure*}
\begin{center}
\subfigure[]{
\includegraphics[angle=270,scale=0.42]{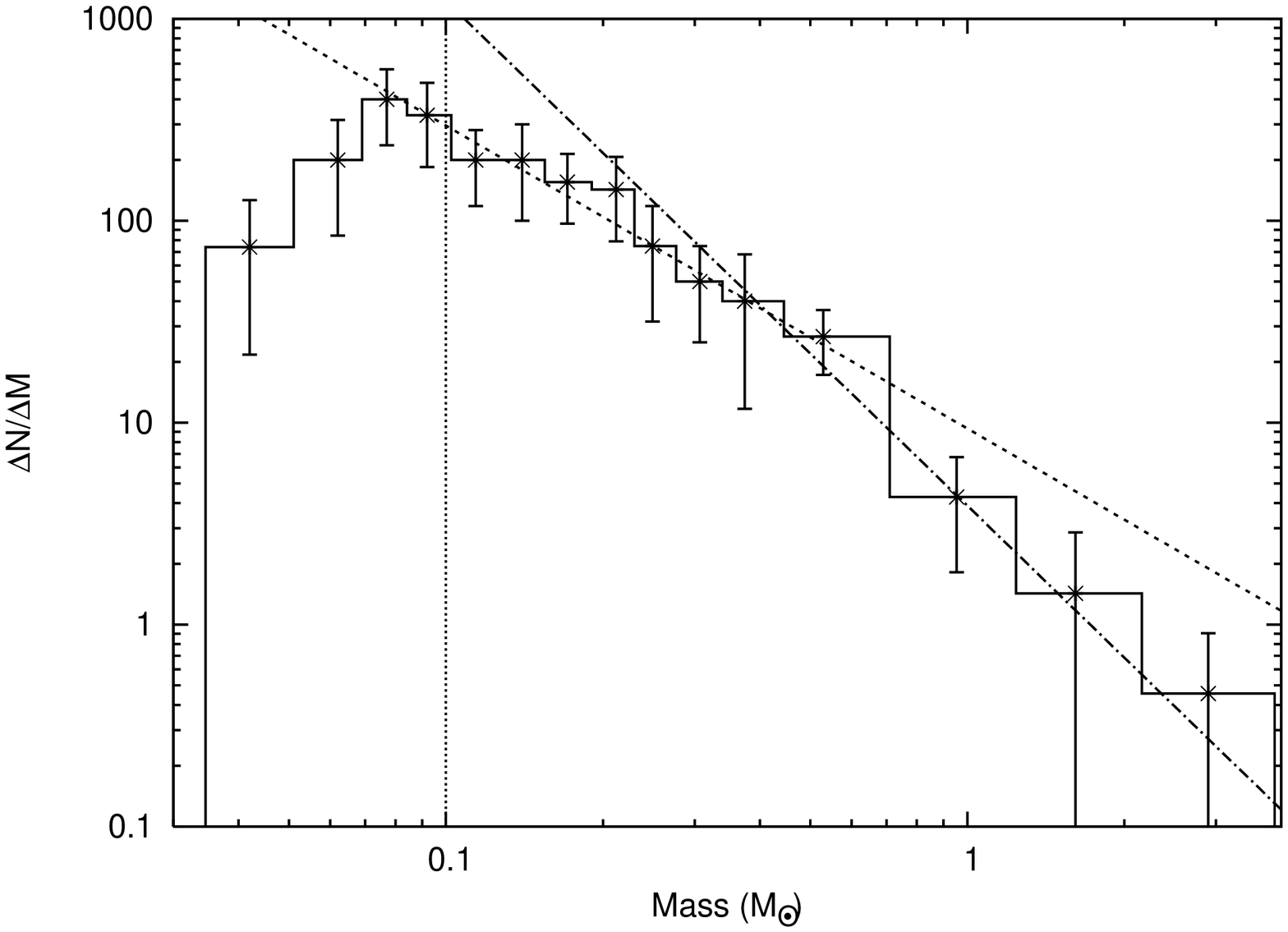}}
\subfigure[]{
\includegraphics[angle=270,scale=0.42]{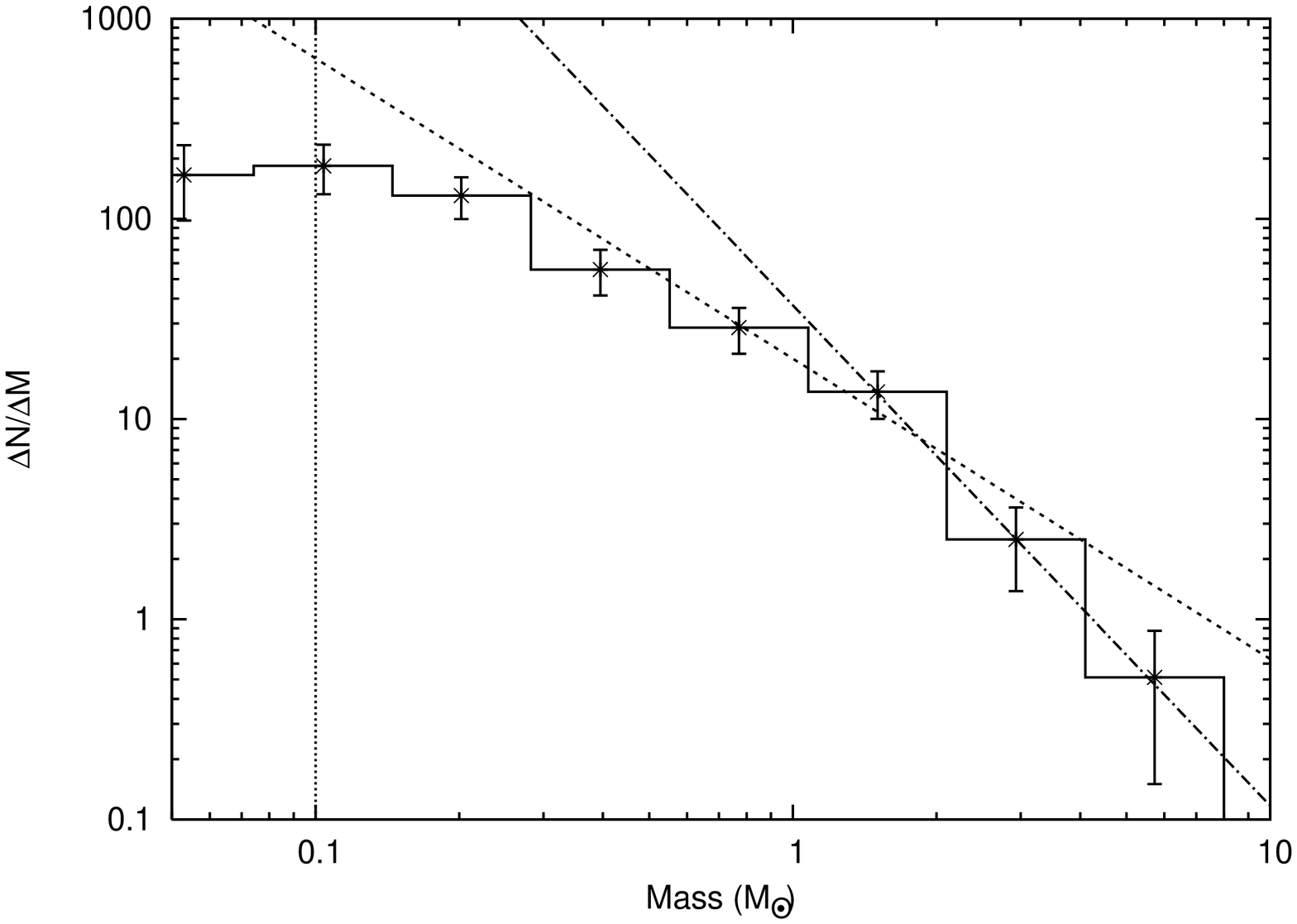}}
\caption{Comparison of 1.3mm and 850$\mu$m datasets. (a) Graph of ($\Delta$N/$\Delta$M) against mass for L1688, recreated from MAN98 1.3mm data for comparison with data from this study. (b) Equivalent graph using SCUBA 850$\mu$m data from JCMT archive. In each plot, the vertical dashed line shows the completeness limit of the study. The two semi-dashed lines on each plot show fits of the two power law slopes from the higher-mass end of the IMF (see text).}
\label{cmf-figs}
\end{center}
\end{figure*}

The 3$\sigma$ contour from the signal-to-noise map was used to draw an elliptical aperture around each identified source on the data. The flux density for any given source was then derived using this 3$\sigma$ aperture and a `sky' aperture of the same size placed on a nearby area in the map containing no significant emission. The locations of each core from each region are given in Table~\ref{Tab1}. Also given in these tables are the equivalent fluxes from MAN98.

The only places on the map where this method for characterizing cores does not apply is when the cores lie on, or adjacent to, bright emission. In these cases, such as cores next to SM1 in Rho Oph-A \citep{1989MNRAS.241..119W}, apertures were drawn following contours on the data map corresponding to the true 3$\sigma$ value and the flux density divided in proportion to the peak fluxes of the cores within areas of bright emission. We found 93 cores, 21 of which had an infrared source associated with them, as identified by Spitzer as part of the c2d surveys \citep{2008arXiv0805.1075E, 2008arXiv0805.0599J}. These latter cores were discounted in our subsequent analysis as not being prestellar.

\citet{2004ApJ...611L..45J} mapped a large area around L1688 using a `fast-mapping' method. This technique is diferent to the standard scan-mapping method and results in lower singal-to-noise data. In order to produce a consistently reduced map using our method, we therefore ignored the non-standard data. However, \citet{2004ApJ...611L..45J} showed that the majority of cores in L1688 are found in the central region, and are covered in this study.

The completeness limit of our data can be estimated using the measured sensitivity of the map, together with the average size of the detected cores. The latter because more mass can be hidden in the noise if the source is larger. In practice this can be done by scaling the integrated flux density from a number of well detected objects, down to the level where the sources would be just undetected, taking into account the selection criteria. For more details of this technique, see \citet{2007MNRAS.374.1413N}. The completeness limit of our map was found to be 0.1 $M_{\sun}$. As an additional check on this limit, we added synthetic sources to the map in a Monte Carlo fashion and attempted to recover them using the method described above. We found that for source sizes up to 20$^{\prime\prime}$ we recovered all of the sources and for sources up to 30$^{\prime\prime}$ we recovered 90\%.

\section{Core Masses}

We assume the 850$\mu$m integrated flux density is optically thin and so the masses of the cores are calculated using the following equation:

\begin{equation}
M=\frac{S_{850}D^2}{\kappa_{850}B_{850,T}},
\label{mass-equ}
\end{equation}

\noindent where $S_{850}$ is the 850~$\mu$m flux density, $D$ is the distance to the source, $\kappa_{850}$ is the mass opacity of the gas and dust, and $B_{850,T}$ is the Planck function at temperature $T$ \citep{2005MNRAS.360.1506K}. Temperatures for each of the regions in the cloud are given in Table~\ref{TabTemp}. We calculate the masses using the revised temperatures given by \citet{2007MNRAS.379.1390S}, hereafter SWW07.

\begin{table}
\caption{Assumed dust temperatures ($T_{dust}$) in K of the regions in the Ophiuchus cloud from MAN98 and SWW07.}
\label{TabTemp}
\centering {
\begin{tabular}{lcc}
\hline

Region & MAN98 & SWW07\\
\hline

Oph-A & 20 & 11\\
Oph-B & 12 & 10\\
Oph-C & 12 & 10\\
Oph-E & 15 & 10\\
Oph-F & 15 & 10\\
Oph-J & --- & 10\\
\hline

\end{tabular}
}
\end{table}

MAN98 assumed a value for dust mass opacity of $\kappa_{1.3mm}$ = 0.005 cm$^{2}$g$^{-1}$ for prestellar cores \citep{1993A&A...279..577P,1996A&A.314.625A}. This is extrapolated to 850$\mu$m using the wavelength dependence factor, $\beta$, which is assumed have a value of 2 in the submillimetre \citep{1983QJRAS..24..267H}. We thus obtain a dust opacity value of $\kappa_{850}$ = 0.01 cm$^{2}$g$^{-1}$.

SWW07 estimated the dust temperatures of clumps in Ophiuchus. They found that in regions where prestellar cores are observed, temperatures of 10-11K are to be expected. Specifically they suggest a dust temperature of 10K for all the main regions identified by this study. The exception to this value is Oph-A, containing SM1 (see Table~\ref{TabTemp}).

For this study we shall assume a distance of 139$\pm$6~pc to the Ophiuchus cloud \citep{2008AN....329...10M}. This value was determined using Hipparcos data and reddening studies \citep{1981A&A....99..346C, 1989A&A...216...44D, 1998A&A...338..897K}. Using these new parameters for distance and temperature a new CMF for the region was created.

\section{Core Mass Function}

In this section we describe CMFs constructed from the above data. We first compare these with previous work, and then see what can be learnt from our new analysis.

\subsection{Comparison with Previous Work}

\subsubsection{MAN98}

Figure~\ref{cmf-figs} shows the data from this study plotted in the same way as the mass spectrum from MAN98, which has been recreated for comparison. Good agreement is seen. Both datasets show the same power law slopes. Below about 2$M_{\sun}$ in our data, the slope in the power law decreases from -1.35 to -0.3 (see section \ref{our-data}). This is in aproximate agreement with the MAN98 results which place this `knee' in the high-mass end of the spectrum at approximately 0.5$M_{\sun}$. The difference in the location of the `knee' is explained by the difference in the temperatures and distances assumed by the two studies (see Table~\ref{TabTemp}).

\begin{figure}
\includegraphics[angle=270,width=0.48\textwidth]{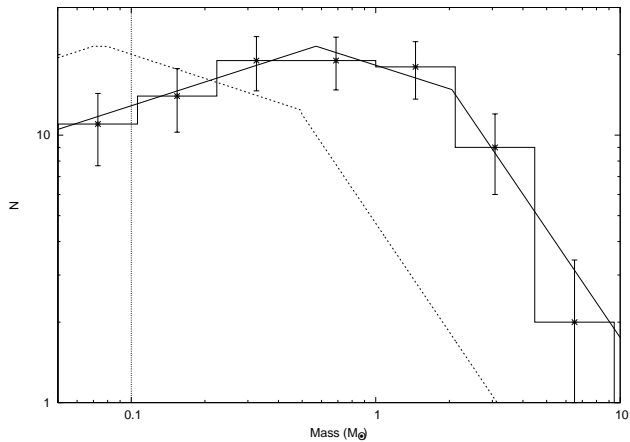}
\caption{CMF for the Ophiuchus dark cloud. The completeness limit is shown at a dashed vertical line at 0.1$M_{\sun}$. A three part stellar IMF (see text), normalised to the peak in N of the CMF, is overlaid as a dotted line. The three power-law slopes of the IMF are also shown fitted to the data as solid black lines, now normalised in M.}
\label{cmf-hist}
\end{figure}

Our mass spectrum also shows another decline in the slope at lower masses in the mass range up to 0.7$M_{\sun}$ (Figure~\ref{cmf-figs}). This effect is seen here, 7 times above our completeness limit, and represents a power law equivalent to the low-mass slope of the stellar IMF.

\subsubsection{Johnstone et al. 2000}

\citet{2000ApJ...545..327J} produced a cumulative CMF for L1688 using some of the same scan-map data (see their Figure~7). Once again, good agreement is seen. Their CMF shows similarities with the two higher-mass slopes of the stellar IMF, in agreement with this study. This earlier study's completeness limit of approximately 0.4$M_{\sun}$ prevents comparison at lower masses.

The CMF produced by \citet{2000ApJ...545..327J} shows a `knee' at around 0.8$M_{\sun}$. As with the MAN98 study, this feature is also seen in our data when an adjustment is made for the different temperature assumptions of the two studies.

\subsubsection{Stanke et al. 2006}

\citet{2006A&A...447..609S} mapped the L1688 region at 1.2mm using IRAM. They produced a mass spectrum with a similar form to that produced in Figure~\ref{cmf-figs}b above. \citet{2006A&A...447..609S} surveyed 111 starless cores. Although this outnumbers the cores identified in this study, many of these are extended, low surface brightness objects that may not be gravitationally bound. They concluded that regardless of the details, the CMF resembles the shape of the IMF with three power-law slopes presented in their CMF plot.

\citet{2006A&A...447..609S} also show the `knee' and a tentative turnover in their CMF. The positions of these feature are not well constrained but their locations at 0.5-1.0M$_{\sun}$ and 0.1-0.3M$_{\sun}$, respectively, are approximately consistent with our data.

\subsubsection{Enoch et al. 2008}

\citet{2008arXiv0805.1075E} surveyed Perseus, Serpens, and Ophiuchus by comparing Bolocam 1.1mm continuum emission maps with Spitzer c2d surveys. Our CMF is consistent with theirs (see their Figure~13).

\subsection{Our Data}
\label{our-data}

The 72 cores in our data with no infrared association are used to plot the CMF in Figure~\ref{cmf-hist}. The dashed vertical line is the 5$\sigma$ completeness limit of 0.1$M_{\sun}$ in our map. This is an improvement over the MAN98 completeness limit, which is due mainly to the better resolution and sensitivity at 850$\mu$m provided by the complete SCUBA data.

Figure~\ref{cmf-hist} shows our CMF. The dotted line shows the stellar IMF, normalised, in N, to the peak of the CMF. The dotted line is a fit to the power laws of the CMF as outlined below. Cores noted to have an infrared source associated with their position in Table~\ref{Tab1} are not included. Compared with Figure~\ref{cmf-figs}, this representation of the data uses a simple number count of cores as opposed to a count divided by the bin size. Figure~\ref{cmf-hist} demonstrates that the data are consistent with the three power-laws in the stellar IMF.

This is the first time that the full turnover in the CMF has been seen in Ophiuchus. Here it is seen at ~0.7M$_{\sun}$. The only other cloud for which this turnover has been detected is Orion \citep{2007MNRAS.374.1413N}. To compare the CMF to the stellar IMF we use a stellar IMF with a power law of the following form:

\begin{equation}
M\frac{dN}{dM} \propto {M}^{-x},
\label{cmf-law}
\end{equation}

\noindent where x takes on different values in different mass regimes. For the stellar IMF we assume the following exponents:

\begin{equation}
\begin{array}{lcrrll}
x & = & 1.35,&   0.5M_{\sun}& < M,& \\
x & = & 0.3,&   0.08M_{\sun}& < M < & 0.5M_{\sun},\\
x &=  & -0.3,&   0.01M_{\sun} & < M < & 0.08M_{\sun}.\\
\end{array}
\label{stellar-imf-x}
\end{equation}

\noindent The higher mass end values for x are taken from \citet{2002Sci...295...82K}. Below 0.08$M_{\sun}$ the IMF is based on fits to young cluster populations used in \citep{2003PASP..115..763C}. In this study values betwen 0.2 and 0.4 are found, so an average of 0.3 is used here.

\begin{table}
\caption{Masses and sizes for cores in the Oph-A and Oph-B regions of the L1688 core. Given the assumption outlined in the text, the error in the masses is $\pm$10\% (see Section~\ref{obs-sec}).}
\label{TabMass1}
\centering{
\begin{tabular}{lcc}
\hline
Core & Mass & Size\\
Name & ($M_{\sun}$) & (AU x AU) \\
\hline

A-MM1 & 0.09 & Unresolved \\
A-MM2/3 & 0.16 & 2800 x 1900 \\
A-MM4 & 0.48 & 2100 x 2100 \\
A-MM5 & 0.82 & 2800 x 2500 \\
A-MM6 & 2.10 & 3600 x 2500 \\
A-MM7 & 0.92 & 3300 x 2200 \\
A-MM8 & 3.18 & 2200 x 1900 \\
A-MM11 & 0.64 & 5600 x 1900 \\
A-MM12 & 0.45 & Unresolved \\
A-MM16 & 0.03 & Unresolved \\
A-MM17 & 0.03 & 1900 x 1900 \\
A-MM18 & 0.87 & 4200 x 2800 \\
A-MM20 & 0.12 & Unresolved \\
A-MM21 & 0.21 & Unresolved \\
A-MM22 & 0.11 & 2800 x 2200 \\
A-MM23 & 1.50 & 4900 x 2100 \\
A-MM26 & 0.43 & 3100 x 1900 \\
A-MM27 & 0.34 & 4200 x 1900 \\
A-MM28 & 0.18 & 2500 x 1900 \\
A-MM29 & 0.16 & Unresolved \\
A-MM30 & 1.28 & 4400 x 1900 \\
A2-MM1 & 0.33 & 3300 x 1900 \\
A3-MM1 & 0.33 & 2500 x 2200 \\
A-N & 0.21 & Unresolved \\
A-S & 0.01 & 2100 x 1900 \\
SM1 & 7.35 & 6100 x 1900 \\
SM1N & 2.91 & 2500 x 1900 \\
SM2 & 5.97 & 5100 x 3600 \\
VLA1623 & 2.93 & 4200 x 2500 \\
\hline
B1-MM1 & 0.02 & Unresolved \\
B1-MM2 & 0.63 & 3100 x 1900 \\
B1-MM3 & 2.61 & 2800 x 2500 \\
B1-MM4 & 2.10 & 5000 x 3200 \\
B1-MM6 & 0.22 & 2500 x 1900 \\
B1-MM7 & 0.11 & 1900 x 1900 \\
B1B2-MM1 & 0.22 & 5600 x 2200 \\
B2-MM2 & 0.88 & 3300 x 2200 \\
B2-MM4 & 0.96 & Unresolved \\
B2-MM5 & 1.14 & 3600 x 2500 \\
B2-MM6 & 1.62 & 5000 x 4200 \\
B2-MM7 & 0.62 & 4200 x 2800 \\
B2-MM8 & 1.63 & 2800 x 2500 \\
B2-MM9 & 1.37 & 3900 x 1900 \\
B2-MM13 & 0.49 & 3100 x 2800 \\
B2-MM14 & 1.26 & 4400 x 2100 \\
B2-MM15 & 0.62 & Unresolved \\
B2-MM16 & 1.31 & 2500 x 2200 \\
\hline

\end{tabular}
}
\end{table}

\begin{table}
\caption{Masses and sizes for cores in the Oph-C, Oph-E, Oph-F and Oph-J regions of the L1688 core. Given the assumption outlined in the text, the error in the masses is $\pm$10\% (see Section~\ref{obs-sec}).}
\label{TabMass2}
\centering {
\begin{tabular}{lcc}
\hline
Core & Mass & Size\\
Name & ($M_{\sun}$) & (AU x AU) \\
\hline

C-MM2 & 1.11 & Unresolved \\
C-MM3 & 1.42 & 4200 x 3300 \\
C-MM5 & 1.24 & Unresolved \\
C-MM6 & 1.05 & 6700 x 3300 \\
C-MM8 & 0.15 & Unresolved \\
C-MM9 & 0.06 & 2200 x 2100 \\
C-MM10 & 0.37 & 2400 x 1900 \\
C-MM12 & 0.09 & 4400 x 2400 \\
C-N & 1.60 & 4700 x 3800 \\
\hline
E-MM2a & 0.12 & 1900 x 1900 \\
E-MM2b & 0.16 & 3100 x 2500 \\
E-MM2d & 0.41 & Unresolved \\
E-MM4 & 0.31 & 3600 x 3100 \\
E-MM5 & 0.28 & 2800 x 2100 \\
E-MM8 & 0.19 & 3100 x 1900 \\
\hline
F-MM1 & 1.05 & 4200 x 2400 \\
F-MM2a & 1.11 & 4400 x 3800 \\
F-MM6 & 0.10 & Unresolved \\
F-MM8 & 0.21 & Unresolved \\
F-MM9 & 0.10 & Unresolved \\
\hline
J-MM2 & 0.42 & 4700 x 3200 \\
J-MM3 & 0.34 & 3900 x 2500 \\
J-MM4 & 0.17 & Unresolved \\
J-MM5 & 0.16 & Unresolved \\
J-MM6 & 0.07 & Unresolved \\

\hline

\end{tabular}
}
\end{table}

\subsection{Star Formation Efficiency}

\begin{figure}
\includegraphics[angle=270,width=0.48\textwidth]{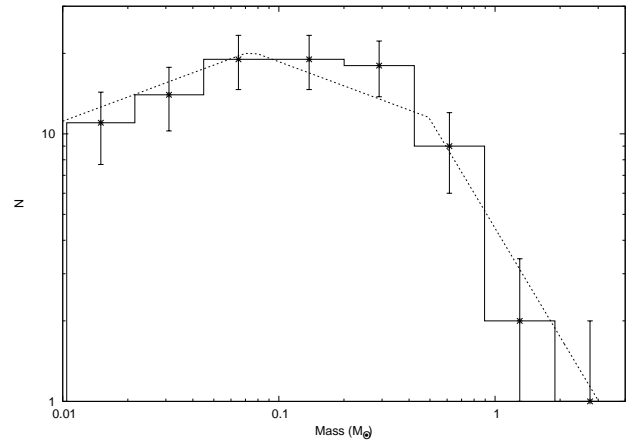}
\caption{CMF for the L1688 core multiplied by a star formation efficiency factor of 0.2, assuming each core forms a single star. A three part stellar IMF, normalised to the peak in N of the CMF, is overlaid as a dotted line (see text for details).}
\label{cmf-SFE}
\end{figure}

\begin{figure}
\includegraphics[angle=270,width=0.48\textwidth]{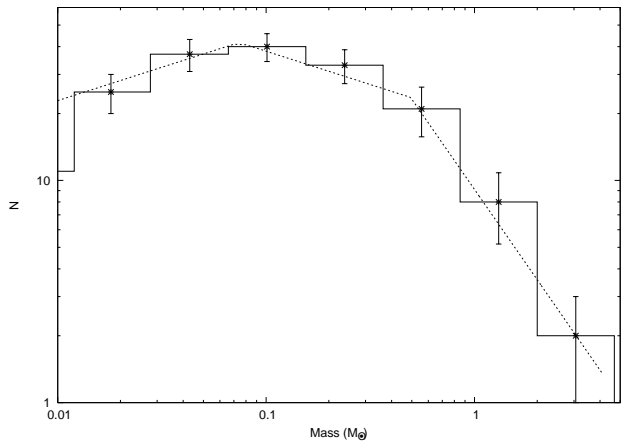}
\caption{Potential stellar IMF for the L1688 core, modelled using the `fully multiple star model' from \citet{2008A&A...477..823G} shown as a histogram. The real stellar IMF is shown as a dotted line. See text for details.}
\label{cmf-goodwin}
\end{figure}

The striking similarity between the CMF and the stellar IMF that is seen in the Ophiuchus L1688 cloud, once again gives a strong indication that the form of the IMF is determined at the prestellar core stage. Furthermore, we have traced this similarity to lower masses than was previously possible. This has allowed us to determine the turnover mass in the core mass function.

If it is assumed that each core will form a single star with a fixed star formation efficiency ($\epsilon$) for all cores, then the CMF can be used to predict an output IMF. In this case the best fit is obtained with $\epsilon$ = 0.2 (see Figure~\ref{cmf-SFE}). This is a simplistic model, and the shape of the CMF does not change as a function of $\epsilon$.

A more sophisticated approach than `one core, one star' would be to model the transformation between core masses and stellar masses in a way that produces multiple stars from each core \citep{2008A&A...482..855H, 2008ApJ...679..552S}. \citet{2008A&A...477..823G} discuss what they call the `fully multiple star model'. This model assumes that cores with masses lower than a certain mass, M$_{CRIT}$, will form binary systems and larger cores will form binary or multiple systems. M$_{CRIT}$ = $\epsilon$$M_{KNEE}$, where $M_{KNEE}$ is the location of the `knee' in Figure~\ref{cmf-hist} (see also \citealt{2004A&A...419..543G}). Thus for this study we use M$_{CRIT}$ = 2$\epsilon$$M_{\sun}$.

Using the `fully multiple star model' it is possible to transform the CMF in Figure~\ref{cmf-hist} to produce a potential, future stellar IMF for L1688 -- see Figure~\ref{cmf-goodwin}. This plot shows very good agreement with the real stellar IMF. The best fit model here requires $\epsilon$ = 0.4.

It is important to remember that this is the star formation efficiency within gravitationally bound cores, and so a high value is to be expected. The total mass of prestellar cores in L1688 is 29.3$M_{\sun}$. The total mass of the L1688 cloud complex is 1447$M_{\sun}$ \citep{1989ApJ...338..902L}. Coupling this to our value of $\epsilon$ = 0.4 for the prestellar cores gives a value for the absolute star formation efficiency of the central portion of the L1688 cloud of about 1-2\%.

\subsection{Comparison with Orion}

\citet{2007MNRAS.374.1413N} used the SCUBA data archive to produce a CMF for the Orion molecular cloud and found a turnover in the CMF at~1.3$M_{\sun}$. This is a factor of two higher than the CMF turnover seen here at~0.7$M_{\sun}$ in the L1688 cloud. However we note that the difference is not significantly greater than one bin-width. Hence, studies with larger number statistics will be needed to verify this hypothesis \citep{2007PASP..119..855W}.

The instrumentation and methodology for deriving the properties of cores is the same in both studies. Both regions can be modelled using the method described in the previous section \citep{2008A&A...477..823G}, but only by using different values for the SFE. These results suggest that the position of the turnover in the CMF may vary with the environment in which the cores reside.

\section{Conclusions}

In this paper we have re-analysed the SCUBA archive data for L1688, incorporating all available high signal-to-noise scan-map and jiggle-map data. An updated form of the CMF in the L1688 cloud complex has been presented using updated values for the distance to this region as well as new estimates for the temperatures of the cores.

We have shown that the CMF for L1688 is consistent with a three part power-law with slopes the same as seen in the stellar IMF. The higher-mass end of the CMF declines as a power law which is consistent with other studies of L1688 \citep[MAN98;][]{2000ApJ...545..327J, 2006A&A...447..609S} as well as studies of Orion \citep{2007MNRAS.374.1413N,2001A&A...372L..41M,2001ApJ...559..307J,2006ApJ...639..259J}.

Hence, the results are mostly in agreement with those found in earlier studies. However, our deeper maps have allowed the discovery of a turnover in the CMF at 0.7$M_{\sun}$ which shows that the core mass function continues to mimic the stellar initial mass function to low masses. This agreement is indicative that the stellar IMF is determined at the prestellar core stage.

It has been shown that the relationship between the CMF and IMF is not necessarily a simple 1:1 translation in the mass axis. Consistency can also be achieved using a fully multiple star model.

\section*{Acknowledgments}

The James Clerk Maxwell Telescope is operated by the Joint Astronomy Centre, Hawaii, on behalf of the UK STFC, the Netherlands NWO, and the Canadian NRC. SCUBA was built at the Royal Observatory, Edinburgh. RJS acknowledges STFC studentship support whilst carrying out this work.

\end{document}